\documentclass[%reprint,
superscriptaddress,twocolumn,
]{revtex4-1}

\usepackage{graphicx}% Include figure files
\usepackage{dcolumn}% Align table columns on decimal point
\usepackage{bm}% bold math

\newcommand{\changed}[1]{#1}

\usepackage{amsmath}
\begin{document}
%\preprint{APS/123-QED}
\title{Distributed quantum sensing in a continuous variable entangled network}

\newcommand{\DTU}{Center for Macroscopic Quantum States (bigQ), Department of Physics, Technical University of Denmark, Fysikvej, 2800 Kgs. Lyngby, Denmark}
\newcommand{\KU}{QMATH, Department of Mathematical Sciences, University of Copenhagen, Universitetsparken 5, 2100 Copenhagen, Denmark}

\author{Xueshi Guo}
\email[E-mail: ]{xguo@fysik.dtu.dk}
\author{Casper R. Breum}%
\affiliation{\DTU}
\author{Johannes Borregaard}
\affiliation{\KU}
\author{Shuro Izumi}
\author{Mikkel V. Larsen}
\affiliation{\DTU}
\author{\changed{Tobias Gehring}}
\affiliation{\DTU}
\author{Matthias Christandl}
\affiliation{\KU}
\author{Jonas S. Neergaard-Nielsen}
\email[E-mail: ]{jsne@fysik.dtu.dk}
\author{Ulrik L. Andersen}
\email[E-mail: ]{ulrik.andersen@fysik.dtu.dk}
\affiliation{\DTU}

\date{\today}

\maketitle
{\bf 
Networking plays a ubiquitous role in quantum technology \cite{Wehner2018}. It is an integral part of quantum communication and has significant potential for upscaling quantum computer technologies \cite{Nickerson2013}. Recently, it was realized that sensing of multiple spatially distributed parameters may also benefit from an entangled quantum network \cite{Komar2014,Eldredge2018,Proctor2018,Ge2018,Zhuang2018,Humphreys2013, DattaPRA2016, Polino2019Optica}. Here we experimentally demonstrate how sensing of an averaged phase shift among four distributed nodes benefits from an entangled quantum network. Using a four-mode entangled continuous variable (CV) state, we demonstrate deterministic quantum phase sensing with a precision beyond what is attainable with separable probes. The techniques behind this result can have direct applications in a number of primitives ranging from molecular tracking to quantum networks of atomic clocks.}

Quantum noise associated with quantum states of light and matter ultimately limits the precision by which measurements can be carried out \cite{Giovannetti2006,Escher2011,Giovannetti2011}. However, by carefully designing the coherence of this quantum noise to exhibit properties such as entanglement and squeezing, it is possible to measure various physical parameters with significantly improved sensitivity compared to classical sensing schemes \cite{CavesPRD1981}. Numerous realizations of quantum sensing utilizing non-classical states of light \cite{Yonezawa2012,Berni2015,Slussarenko2017} and matter \cite{Muessel2014} have been reported, while only a few applications have been explored. Examples are quantum-enhanced gravitational wave detection \cite{LIGO}, detection of magnetic fields \cite{Wolfgramm2010,Li2018,Jones2009} and sensing of the viscous-elasticity parameter of yeast cells \cite{Taylor2013}. All these implementations are, however, restricted to the sensing of a single parameter at a single location.

Spatially distributed sensing of parameters at multiple locations in a network is relevant for applications from local beam tracking~\cite{Qi2018} to global scale clock synchronization~\cite{Komar2014}. 
The development of quantum networks enables new strategies for enhanced performance in such scenarios. Theoretical works \cite{Humphreys2013,Knott2016,Baumgratz2016,Pezze2017,Eldredge2018,Proctor2018,Ge2018,Zhuang2018} have shown that entanglement can improve sensing capabilities in a network using either twin-photons or Greenberger-Horne-Zeilinger (GHZ) states combined with photon number resolving detectors \cite{Proctor2018,Ge2018} or using CV entanglement for the detection of distributed phase space displacements \cite{Zhuang2018}. 
In this Letter, we experimentally demonstrate an entangled CV network for sensing  the average of multiple phase shifts inspired by the theoretical proposal of Ref.~\cite{Zhuang2018}. We focus on the task of estimating small variations around a known phase in contrast to \emph{ab initio} phase estimation. For the first time in any system, we demonstrate deterministic distributed sensing in a network of four nodes with a sensitivity beyond that achievable with a separable approach using similar quantum states.

\begin{figure}[ht]
\centering
\includegraphics[width=\linewidth]{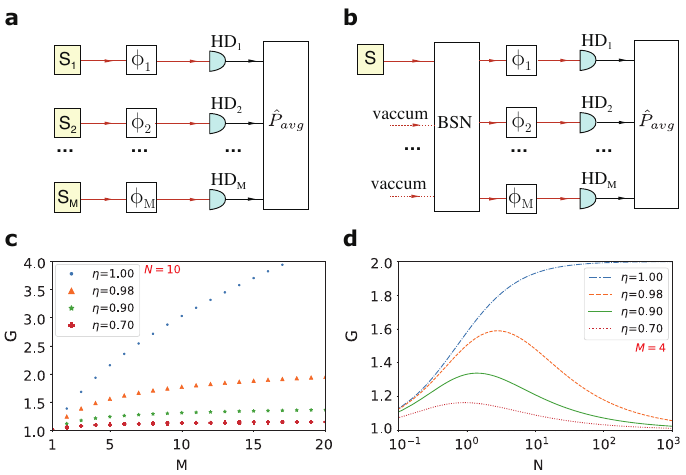}
\caption{
\textbf{Distributed phase sensing scheme}. The task is to estimate the average value of $M$ spatially distributed small phase shifts $\phi_1,\ldots,\phi_M$. (\textbf{a}) Without a network, the average phase shift must be estimated by probing each sample individually. This can be done with homodyne detection of the phase quadrature (HD$_1$,$\ldots$,HD$_M$), and the sensitivity can be increased by using squeezed probes  generated by $M$ independent squeezers $S_1,\ldots, S_M$. 
(\textbf{b}) If the $M$ sites are connected by an optical beam splitter network (BSN), a single squeezed probe can be distributed among the sites. This enables entanglement-enhanced sensing of the average phase shift.
(\textbf{c,d}) The entangled approach of panel (b) shows a gain in sensitivity compared to the separable approach in panel (a) for the same number of photons, $N$, hitting each sample and with optimized probe states.
This gain, $G=\sigma^\mathrm{opt}_{s}/\sigma^\mathrm{opt}_{e}$, is here plotted as a function of the number of samples $M$ with $N$ fixed at 10 (c) and as a function of the average number of photons with $M$ fixed at 4 (d) for different values of $\eta$, the efficiency of the channel between pure resource state and phase sample. 
}
\label{fig_theory}
\end{figure}
  
We start by introducing a theoretical analysis of the networked sensing scheme assuming the existence of an external phase reference. Consider a network of $M$ nodes with optical inputs that undergo individual phase shifts, $\phi_j\ (j=1, \dots, M)$. The goal is to estimate the averaged phase shift, $\phi_\mathrm{avg}=\sum_{j=1}^M\phi_j/M$, among all nodes with as high precision as possible. Two different sensing setups are considered: A separable system where the nodes are interrogated with independent quantum states (Figure \ref{fig_theory}a) and an entangled system where they are interrogated with a joint quantum state (Figure \ref{fig_theory}b). 
We assume the squeezers give out pure single-mode Gaussian quantum states described by the state vectors $\hat D(\alpha)\hat S(r)|0\rangle$, where $\hat D$ and $\hat S$ are the displacement and squeezing operators, respectively, $\alpha$ is the displacement amplitude and $r$ is the squeezing factor. We assume that each probe state undergoes loss in a channel with transmission $\eta$.
We furthermore restrict the estimator to be the joint phase quadrature, $\hat P_\mathrm{avg}=\sum_{j=1}^M \hat p_j/M$ (where $\hat{p}_j$ are the phase quadratures of the individual modes), practically corresponding to the averaged outcome of $M$ individual homodyne detectors. These states and detectors are of particular interest due to their experimental feasibility, inherent deterministic nature, high efficiency, and robustness to noise.

Using the separable approach, $M$ %different 
identical Gaussian probe states are prepared and individually detected, while in the entangled approach, a single squeezed Gaussian state is distributed evenly to the $M$ nodes via a beam splitter array and likewise measured individually with homodyne detectors at the nodes. 
If one wanted to estimate different linear combinations of the phase shifts than the simple average, other beam splitter divisions would be required \cite{Eldredge2018,Proctor2018}.
The sensitivity of the measurement can be defined as the standard deviation of the measurement which, by error propagation, is \cite{Giovannetti2011}
\begin{equation}
\label{Eq_sensitivity_def}
  \sigma= \frac{\sqrt{\langle\Delta \hat P^2_\mathrm{avg} \rangle}}{|\partial \langle \hat P_\mathrm{avg}\rangle /\partial\phi_\mathrm{avg}|},  
\end{equation}
where $\langle \Delta \hat P^2_\mathrm{avg} \rangle=\langle\hat P^2_\mathrm{avg}\rangle-\langle\hat P_\mathrm{avg}\rangle^2$ is the variance of the estimator.
We are only interested in the sensitivity for small phase shifts, since one can always use an initial rough phase estimation to adjust the homodyne detector (the local oscillator phase) to the maximum sensitivity setting \cite{Berni2015}. For small phase shifts, we obtain the sensitivities for the separable ($\sigma_s$) and entangled ($\sigma_e$) approaches (see Supplementary Material Sec. I):
\begin{align}
\label{eq:sens_sep}
\sigma_s &= \frac{\sqrt{e^{-2 r_s} + 1/\eta - 1}}{2 \alpha_s \sqrt{M}} ,\\
\sigma_e &= \frac{\sqrt{e^{-2 r_e} + 1/\eta - 1}}{2 \alpha_e} .
\end{align}
We now constrain the average number of photons, $N$, hitting each sample. 
The photons can be separated into those originating from coherent displacement and those originating from squeezing:
$N=N_{s,\mathrm{coh}} + N_{s,\mathrm{sqz}} = \eta(\alpha_s^2 + \sinh^2 r_s)$ for the separable case and $N=N_{e,\mathrm{coh}} + N_{e,\mathrm{sqz}} = \eta(\alpha_e^2 + \sinh^2 r_e)/M$ for the entangled case.
The ratio between photon numbers, parametrized as $\mu_{s(e)} = N_{s(e),\mathrm{sqz}} / N$ can be tuned to give the optimal sensitivities
\begin{align}
\sigma_s^\mathrm{opt} &= \frac{1}{2\sqrt{M}N} \sqrt{\frac{N(1-\eta) + \frac{\eta}{2} \big(1 + \sqrt{1+4N(1-\eta)} \big)}{1 + \eta/N}} , \\
\sigma_e^\mathrm{opt} &= \frac{1}{2MN}\sqrt{\frac{MN(1-\eta) + \frac{\eta}{2} \big(1 + \sqrt{1+4MN(1-\eta)} \big)}{1 + \eta/(MN)}} .
\end{align}

For perfect efficiency ($\eta=1$), it is clear that the sensitivity of the entangled system yields Heisenberg scaling both in the number of nodes $(1/M)$ and the number of photons per mode ($1/N$) whereas the separable system only achieves the latter and a classical $1/\sqrt{M}$-scaling with the number of modes. The gain in sensitivity of the entangled network relative to the separable network (denoted $G=\sigma^\mathrm{opt}_{s}/\sigma^\mathrm{opt}_{e}$) is thus $G=\sqrt{M}$.
 
For non-ideal efficiency, the Heisenberg scaling ceases to exist  in accordance with previous work on single parameter estimation~\cite{KnyshPRA2011}. In fact, for $\eta\rightarrow 0$, both sensitivities approach $1/2\sqrt{MN}$. Still, it is important to note that the entangled network exhibits superior behavior for any value of $\eta$, $M$ and $N$ for optimized $\mu_s,\mu_e$. 
Some examples for the sensitivity gain are illustrated in Figures \ref{fig_theory}c and d.
From Fig \ref{fig_theory}d where a network of $M=4$ nodes is considered, it is clear that the highest gain in sensitivity is attained at a finite photon number. 
We also note that for large photon numbers, the gain tends to unity for non-zero loss meaning no enhanced sensitivity when using the entangled approach.  However, there is still a practical advantage for the entangled approach: Only one squeezed state is needed compared to the M squeezed states with similar squeezing levels for the separable approach (see Supplementary Material Sec. I).

\begin{figure*}[!ht]
\centering
\includegraphics[width=0.8\linewidth]{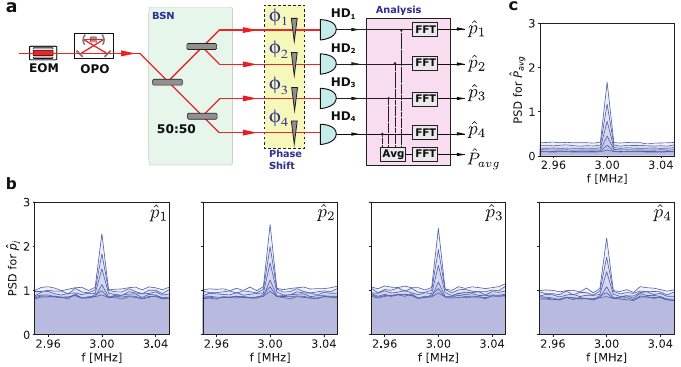}
\caption{\textbf{Experimental setup and data.} \textbf{(a)} A schematic outline of the experimental setup for the entangled approach with $M=4$ (see Supplementary Material Sec. III for more details).
A 1550 nm laser beam is phase-modulated at 3 MHz by an electro-optic modulator (EOM) and injected into an optical parametric oscillator (OPO).
This prepares a displaced squeezed state at the 3 MHz side-band.
A beam-splitter network (BSN) splits the state into four identical and entangled probes which are used to sense the average phase shift of $\phi_1$ to $\phi_4$ introduced by four $\lambda/2$ wave-plates. After phase shifting, the probes' phase quadratures $\hat{p}_j$ are measured with homodyne detection setups (HD$_1$ to HD$_4$) whose outputs are recorded by an oscilloscope. 
The power spectral densities (PSD) of the individual modes as well as the average of them, $\hat{P}_\mathrm{avg}$, are obtained from Fourier transformations (FFT) of the oscilloscope traces. 
This setup can be reduced to the separable approach ($M=1$) by removing the BSN and sending the state to one phase shift and HD.
\textbf{(b,c)} PSD results,  estimated from 2000 FFT measurement with 5 kHz resolution, for $\hat p_{j}$ \textbf{(b)} and $\hat P_\mathrm{avg}$ \textbf{(c)} from one experimental run for the entangled approach $M=4$. We rotate all the $\lambda/2$ wave-plates by 1$^\circ$ and measure the side-band spectra for different $\phi_\mathrm{avg}$, known through phase calibration (Supplementary Sec. IV).  The values of $\phi_\mathrm{avg}$ are $0.3^\circ$, $4.2^\circ$,  $8.2^\circ$ ,  $12.1^\circ$,  $16.1^\circ$,  $20.0^\circ$ from the bottom to the top curve in the plots. Due to the quantum correlations of the entangled probes, the noise of $\hat P_\mathrm{avg}$ reduces significantly compared to $\hat p_{j}$. 
From these spectra, $\langle \hat P_\mathrm{avg} \rangle$ and $\langle \Delta \hat P_\mathrm{avg}^2 \rangle$, which constitute the sensitivity $\sigma$, are extracted: The peak and the noise level of the spectrum for $\hat{P}_\mathrm{avg}$ are respectively given by $ \langle \hat P_\mathrm{avg}^2 \rangle $ and $\langle \Delta \hat P_\mathrm{avg}^2 \rangle = \langle \hat P_\mathrm{avg}^2 \rangle -\langle \hat P_\mathrm{avg} \rangle^2 $. 
}
\label{fig_setup}
\end{figure*}

Next, we demonstrate experimentally the superiority of using an entangled network for distributed sensing. The entangled network is realized by dividing equally a displaced single mode squeezed state into four spatial modes by means of three balanced beam splitters (Fig. \ref{fig_setup}a, see Supplementary Material Sec. III for more details). These modes are then sent to the four nodes of the network at which they each undergo a phase shift $\phi_j$ and are finally measured with high-efficiency homodyne detectors (HD) that are set to measure the phase quadrature, $\hat{p}_j$. The external phase reference is set by the local oscillator which co-propagates with the probes through the setup but in a different polarization mode. This ensures that the relative phases between the probes and the local oscillator can be controlled. The resulting photo-currents from the four detectors are further processed and subsequently combined to produce the averaged phase shift.
For demonstration purpose, we set all $\phi_j$ to the same value $\phi_j=\phi_{avg}$, but in principle they could be different. 

We choose to define our quantum states within a narrow spectral mode at the 3 MHz sideband frequency. There are no fundamental restrictions in the scheme on the optical modes employed. In any practical setting, they would be chosen based on the nature of both the squeezing source and the samples being probed. Here, the 3 MHz sideband is chosen to maximize the squeezing from our source, an optical parametric oscillator (OPO) operating below threshold: At higher frequencies, the squeezing reduces due to the limited bandwidth of the OPO, while at lower frequencies, it is degraded by technical noise.
A displaced squeezed state is obtained by injecting into the OPO a coherent state produced by phase modulating the injected beam at 3 MHz.
The maximum squeezing measured through the joint measurement of 4 HDs is $\sim$5 dB at 3 MHz. 
More details on the probe generation are in Supplementary Material Sec. III.

An experimental run is shown in Fig. \ref{fig_setup}b. In this particular run, a displaced squeezed state with an average photon number of $N=2.48\pm0.12$ in each mode is prepared of which $N_{e,\mathrm{sqz}}=0.30\pm0.01$ photons are from the squeezing operation and $N_{e,\mathrm{coh}}=2.19\pm0.11$ are from the phase modulation as this distribution is near-optimal for the entangled case. We then impose 12 different $\phi_\mathrm{avg}$ values by phase shifts at each node while recording the Fourier transformed homodyne detector outputs; the spectra around the 3 MHz sideband for six of the $\phi_\mathrm{avg}$ values are shown in Fig.~\ref{fig_setup}b (see Supplementary Material Sec. V for more details). 
These outputs yield poor estimates of the individual phase shifts (because the squeezing in each mode is only $\sim$0.8 dB) but the averaged phase shift obtained by summing the photo-currents produces an entanglement-enhanced estimate with significantly lower noise. The spectra for the averaged photo-currents are shown in Figure \ref{fig_setup}c.
For comparison, we also simulate the separable approach by directing the entire displaced squeezed state (with properly optimized parameters) to a single node. We then perform the phase estimation at that node and scale the obtained sensitivity by $\sqrt{4}$ to get the projected performance for an average over four identical sites.
An example is shown in Fig \ref{fig_setup}b for $N=2.63\pm0.11$, with $N_{s,\mathrm{sqz}}=0.31\pm0.01$ and $N_{s,\mathrm{coh}}=2.32\pm0.10$.   

\begin{figure*}[ht]
\centering
\includegraphics[width=.8\linewidth]{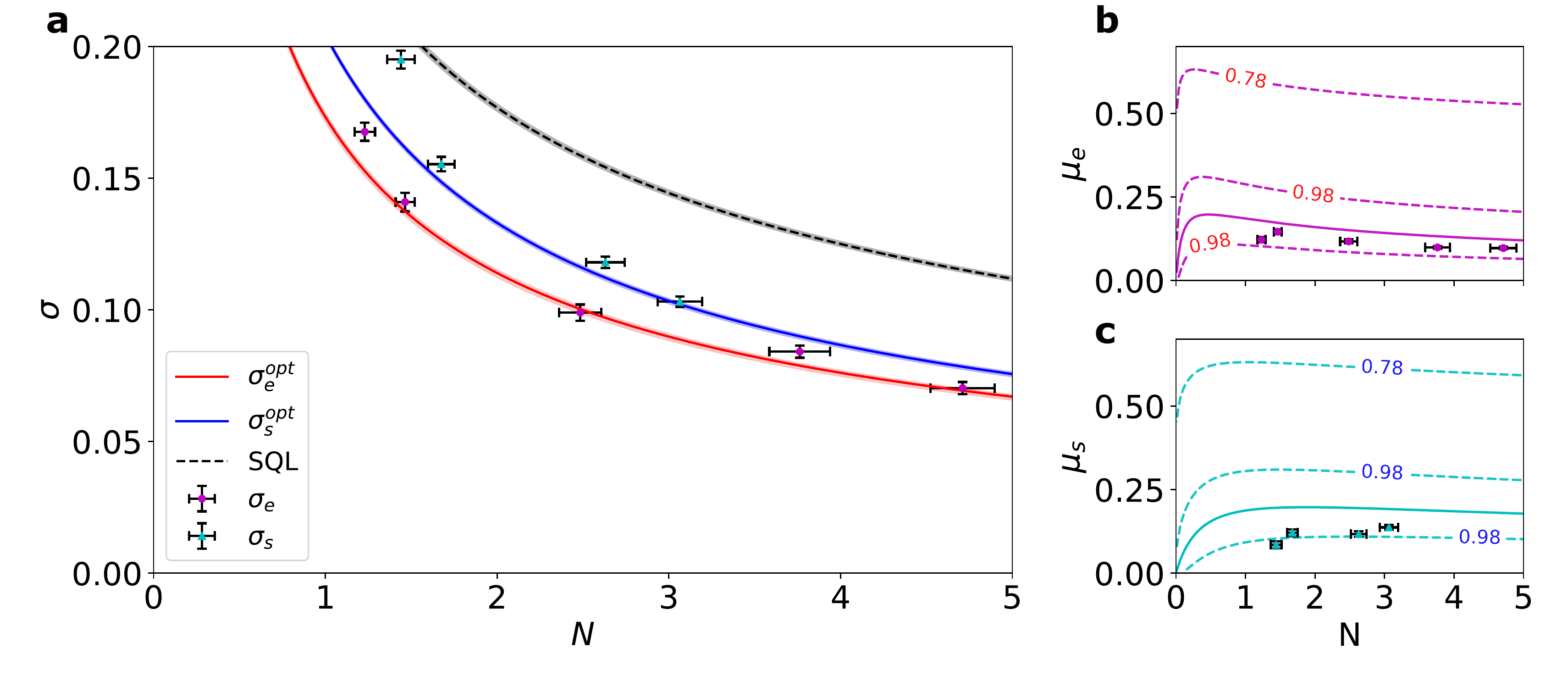}
\caption{\textbf{Phase sensitivity results.} \textbf{(a)} Sensitivity to $\phi_\mathrm{avg}$ for different average number of photons per sample $N$ for the entangled scheme ($\sigma_e$) and the separable scheme ($\sigma_s$).  The sensitivities predicted in theory, $\sigma_e^\mathrm{opt}/\sigma_s^\mathrm{opt}$, are plotted with shadowed lines, where the shadows show the upper/lower bound within the overall  efficiency $\eta=73.5\%\pm1.5\%$ of our experimental setup. SQL: the standard quantum limit, for which no squeezer but only coherent states are used. 
The result shows that both schemes perform better than the SQL and that the entangled network outperforms the separable network.
\textbf{(b)} Data points are the values of $\mu_e$ (the proportion of $N$ originating from the squeezing process) obtained in the experiment.
The solid curves are the optimal $\mu_e$ that minimize $\sigma$ at a given $N$. 
The contours indicate the values of $\sigma_e^\mathrm{opt}/\sigma_e$.
\textbf{(c)} As \textbf{(b)}, but for the separable approach. All error bars are plus/minus standard deviation assuming normal distribution.
}
\label{fig_result}
\end{figure*}

We quantify the performance of the sensing network by estimating the sensitivities of the two approaches based on the averaged homodyne measurement outcomes, $P_\mathrm{avg}$. By extracting the rate of change with respect to a phase rotation, $|\partial\langle \hat P_\mathrm{avg}\rangle / \partial\phi_\mathrm{avg}|$, as well as the variance, $\langle \Delta \hat{P}_\mathrm{avg}^2\rangle$, of $P_\mathrm{avg}$, we deduce the sensitivity using Eq. (\ref{Eq_sensitivity_def}). 
For the experimental runs described above, we obtain sensitivities of $\sigma_{e} = 0.099 \pm 0.003$ and $\sigma_{s} = 0.118 \pm 0.002$ for the entangled and separable approach, respectively. This corresponds to single shot resolvable distributed phase shifts (that is, phase shifts for which the signal-to-noise ratio is unity) of $5.66^\circ \pm 0.18^\circ$ for the entangled case and $6.76^\circ \pm 0.11^\circ$ for the separable case with $\sim 2.5$ photons. Using a coherent state in replacement of the squeezed state, the minimal resolvable phase for $2.5$ photons is $9.06^\circ \pm 0.07^\circ$ corresponding to the standard quantum limit. 
Note that these angles are larger than our small phase shift approximation (which requires $\phi_{avg}$ to be much smaller than $\sim7^{\circ}$ for the conditions in this experimental run, see Supplementary Sec. I). 
In practice this means that it is necessary to probe the sample more than once to resolve the small phases implemented in the experiment. Sampling the phases $K$ times will result in $\sqrt{K}$ times smaller resolvable phase shift angles. The entangled strategy will still benefit from the enhanced sensitivity per probe. 

We find the sensitivities for different total average photon numbers both for the entangled and separable network, and plot the results in Figure \ref{fig_result}a. For every selection of the total photon number, we adjust $\mu$ to a near-optimal value for optimized sensitivity (Figure \ref{fig_result}b,c). It is clear in Figure \ref{fig_result}a that both realizations beat the standard quantum limit (reachable by coherent states of light), and most importantly, we see that the entangled network outperforms the separable network. 
The ultimate sensitivity of our entangled approach is not reached in our implementation. However, homodyne detection will not even in principle saturate this bound and non-Gaussian measurements are in general needed (see Supplementary Material Sec. II).

Our results experimentally demonstrate how mode entanglement,  here in the form of squeezing of a collective quadrature of a multi-mode light field, can enhance the sensitivity in a distributed sensing scenario. Importantly, we show this enhancement in an experimentally feasible setting where the sensitivity of standard coherent probes are enhanced through quadrature squeezing. This approach allows for easily tunable probe powers in order to adapt the setup to the specific application. \changed{ Furthermore, because the entanglement is generated from a simple beam-splitter network, it is straight-forward to scale to more modes where the sensitivity gain may be even larger, cf. Fig.~\ref{fig_theory}c. The main limitation will be the channel efficiency which will eventually limit the gain.}  Consequently, we believe that techniques demonstrated here have direct applications in a number of areas. Specifically, beam tracking relevant for  molecular tracking~\cite{Taylor2013,Qi2018} could directly benefit from these techniques. 
Such applications impose limits on the allowed probe power to prevent photon damage and heating of the systems. Mode-entanglement can thus be used to increase sensitivity without increasing the probe power. Using squeezed coherent light for quantum non-demolition (QND) measurement has also been exploited for generation of spin squeezing in atomic ensembles~\cite{Hammerer2010} and optical magnetometry~\cite{Wolfgramm2010}. While this is usually considered for single ensembles, the generalization to multiple ensembles can provide enhanced sensitivity and new primitives for quantum information processing. Combining several ensembles for magnetometry and utilizing mode-entanglement would further reduce the shot-noise and increase sensitivity of a collective optical measurement. Performing a collective optical QND measurement of several atomic ensembles can prepare a distributed spin-squeezed state for quantum network applications. In particular, squeezing of multiple optical lattice clocks could be used for collective enhancement of clock stability~\cite{Komar2014,Eugene2016}. In Ref.~\cite{Eugene2016}, this was obtained by letting a single probe interact with all ensembles in a sequential manner. However, utilizing mode-entanglement, this can be performed in a parallel fashion with no quantum signal being transmitted between the ensembles. 

\vspace{7mm}\noindent\textbf{DATA AVAILABILITY}\vspace{2mm}\\
Experimental data and analysis code is available on request.

\vspace{7mm}\noindent\textbf{Acknowledgments}\vspace{2mm}\\
M.C. and J.B. acknowledge support from VILLUM FONDEN via the QMATH Centre of Excellence (Grant no. 10059), the European Research Council (ERC Grant Agreements no 337603), and from the QuantERA ERA-NET Cofund in Quantum Technologies implemented within the European Union's Horizon 2020 Programme (QuantAlgo project) via the Innovation Fund Denmark. 
X.G., C.B., S.I., M.L., T.G., J.N. and U.A. acknowledge support from Center for Macroscopic Quantum States (bigQ DNRF142). X.G., S.I. and J.N. acknowledge support from VILLUM FONDEN via the Young Investigator Programme (Grant no. 10119).

\vspace{7mm}\noindent\textbf{AUTHOR CONTRIBUTIONS}\vspace{2mm}\\
J.B., U.A., J.N., T.G. and X.G. conceived the experiment. X.G., C.B. and M.L. performed the experiment and analyzed the data. J.B., X.G., S.I., M.C. and J.N. worked on the theoretical analysis. 
X.G. wrote the paper with contributions from J.B., C.B., S.I., J.N. and U.A.
J.N. and U.A. supervised the project.

\vspace{7mm}\noindent\textbf{ADDITIONAL INFORMATION}\vspace{2mm}\\
\textbf{Competing interests:} The authors declare that there are no competing interests.


%%%%%%%%%% Merge with supplemental materials %%%%%%%%%%
\pagebreak
\widetext
\begin{center}
\textbf{\large Supplemental Materials for Distributed quantum sensing in a continuous variable entangled network}
\end{center}

\setcounter{equation}{0}
\setcounter{figure}{0}
\setcounter{table}{0}
\setcounter{page}{1}
\makeatletter
\renewcommand{\theequation}{S\arabic{equation}}
\renewcommand{\thefigure}{S\arabic{figure}}
\renewcommand{\bibnumfmt}[1]{[S#1]}
\renewcommand{\citenumfont}[1]{S#1}

\section{Averaged phase shift sensing with
$\hat P_\mathrm{avg}$ estimator}\label{sec:sensitivity}

Our distributed phase sensing scenario is as follows. At each of \(M\)
spatially separated locations, an optical phase shift \(\phi_j\) occurs.
We are interested in estimating the average phase shift
\(\phi_\mathrm{avg} = \frac{1}{M} \sum_{j=1}^{M}\phi_j\). It is
straight-forward to generalize to other linear combinations of the phase
shifts, but for the sake of demonstrating the power of the entangled
approach it suffices to consider the simple average, where the gain is maximum \cite{ProctorPRL2018}. We consider two
different approaches: The \emph{separable} scheme where each phase shift
is probed individually by squeezed coherent states, and the
\emph{entangled} scheme where the \(M\) locations are part of an optical
network endowed with a single squeezed coherent state that is
distributed among the nodes to serve as an entangled probe. In either
case, the phase shifted probes are measured by homodyne detection of
their phase quadratures and the results are communicated classically to
establish the average.

We furthermore make the following assumptions to simplify the analysis:

\begin{enumerate}
\def\labelenumi{\arabic{enumi}.}
\item
  All the phase shifts are small, giving the small-angle approximation
  \(\sin\phi \approx \phi\).
\item
  All probes in the separable approach are identical, having real-valued
  displacement amplitude \(\alpha_s\) and squeezing in the phase
  quadrature with squeezing parameter \(r_s\). That is, the \(M\)
  probes are each in the state
  \(|\psi^{(s)}\rangle = \hat{D}(\alpha_s) \hat{S}(r_s) |0\rangle\), where \(\hat{D}(\alpha) = \exp(\alpha \hat{a}^\dag - \alpha^\ast \hat{a})\) is the displacement operator and \(\hat{S}(r) = \exp(\frac{r}{2}(\hat{a}^{\dag 2} - \hat{a}^2)\)) is the squeezing operator.
\item
  In the entangled approach, the single initial resource state has
  real-valued displacement amplitude \(\alpha_e\) and phase squeezing with
  squeezing parameter \(r_e\), that is, it is in the state
  \(|\psi^{(e)}\rangle = \hat{D}(\alpha_e) \hat{S}(r_e) |0\rangle\). This
  resource is divided evenly through the network to the \(M\) nodes.
\item
  The channel losses, quantified by the efficiency parameter \(\eta\),
  are identical for the \(M\) channels and they occur entirely prior to
  the probes reaching the phase samples. In other words, we assume the
  phase samples themselves and the detection to be lossless. While this
  assumption is not quite realistic, even in our experiment, it mostly
  has consequences when keeping track of the number of photons hitting
  the sample but does not influence the sensitivity as such. In a truly
  distributed setting, most losses would also happen in the distribution
  of the resources.
\end{enumerate}

For high-sensitivity estimation of larger phase shifts, these
assumptions can still be fulfilled, as long as the local oscillator in
the homodyne detector is pre-adjusted to be roughly 90$^\circ$ out of phase
with the shifted probe. This rough estimation can be done with just a
few initial probings \cite{Berni2015}.

\subsection{General sensitivity for small phase shift}\label{header-n14}

\subsubsection{Separable scheme}\label{header-n73}

\begin{figure}[h]
\centering
\includegraphics[width=0.3\linewidth]{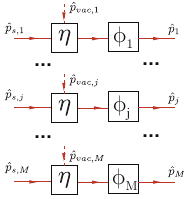}
\caption{Phase quadrature notations for analyzing the separable scheme. The amplitude quadrature is defined accordingly. $\hat p_{s,1}\dots\hat p_{s,M}$: operator for the initial squeezed states; $\hat p_{\mathrm{vac},1}\dots\hat p_{\mathrm{vac},M}$: the vacuum operators induced by loss. $\eta$: the overall detection efficiency; $\phi_1\dots\phi_M$: the local phase shifts; $\hat p_{1}\dots\hat p_{M}$: phase quadrature of a single mode after channel loss and the phase shift.}
\label{fig:sep}
\end{figure}

With probe states given as described above, we use the notation defined in Figure \ref{fig:sep} to analyse the separable scheme. The phase quadrature of a
single mode after channel loss and the phase shift \(\phi_j\) is

\begin{equation}
\hat{p}_j = \left( \sqrt{\eta}\,\hat{x}_{s,j} + \sqrt{1-\eta}\,\hat{x}_{\mathrm{vac},j} \right) \sin\phi_j  +  \left( \sqrt{\eta}\,\hat{p}_{s,j} + \sqrt{1-\eta}\,\hat{p}_{\mathrm{vac},j} \right) \cos\phi_j,
\end{equation}
where
\(\hat{x}_{s,j}, \hat{p}_{s,j}\) are
the quadrature operators of the initial squeezed states with mean values
\(\langle \hat{x}_{s,j} \rangle = \sqrt{2}\alpha_s\),
\(\langle \hat{p}_{s,j} \rangle = 0\) and variances
\(\langle \Delta \hat{x}_{s,j}^2 \rangle = \frac{1}{2} e^{2r_s}\),
\(\langle \Delta \hat{p}_{s,j}^2 \rangle = \frac{1}{2} e^{-2r_s}\),
while \(\hat{x}_{\mathrm{vac},j}\), \(\hat{p}_{\mathrm{vac},j}\) are
vacuum mode operators admixed through the losses. The expectation value
of the rotated phase quadrature is

\begin{equation}
\langle \hat{p}_j \rangle = \sqrt{\eta}\, \langle \hat{x}_{s,j} \rangle \sin\phi_j = \sqrt{2 \eta} \alpha_s \sin\phi_j \approx \sqrt{2\eta} \alpha_s \phi_j.
\end{equation}

The phase shift can thus be directly estimated from the measured
\(\hat{p}_j\) values. The average phase shift of \(M\) modes,
\(\phi_\mathrm{avg} = \frac{1}{M} \sum_{j=1}^{M}\phi_j\), can then be
estimated with the estimator
\(\hat{P}_\mathrm{avg} = \frac{1}{M} \sum_{j=1}^M \hat{p}_j\):

\begin{equation}
\langle \hat{P}_\mathrm{avg}\rangle \approx \sqrt{2\eta}\alpha_s \phi_\mathrm{avg} .
\end{equation}

The sensitivity of the estimation is defined as the standard deviation
which, from standard error propagation analysis, is given by

\begin{equation}
\label{eq:sens_def}
\sigma_s = \frac{\sqrt{\langle \Delta \hat{P}_\mathrm{avg}^2 \rangle}}
                    {\left| \partial \langle \hat{P}_\mathrm{avg} \rangle / \partial \phi_\mathrm{avg} \right|} .
                    \end{equation}

The slope of \(\hat{P}_\mathrm{avg}\) versus \(\phi_\mathrm{avg}\) is

\begin{equation}
\partial \langle \hat{P}_\mathrm{avg} \rangle / \partial \phi_\mathrm{avg} \approx \sqrt{2 \eta} \alpha_s,
\end{equation}
and its variance is

\begin{align}
\langle \Delta \hat{P}_\mathrm{avg}^2 \rangle &= 
\frac{1}{M^2} \left\langle \Delta \left(\sum_{j=1}^M \hat{p}_j \right)^2 \right\rangle = 
\frac{1}{M^2} \sum_{j=1}^M \langle \Delta\hat{p}_j^2 \rangle \\ &= 
\frac{1}{M^2} \sum_{j=1}^M \left( 
\sin^2\!\phi_j (\eta \langle\Delta\hat{x}^2_{s,j}\rangle + 
                (1-\eta) \langle\Delta\hat{x}^2_{\mathrm{vac},j}\rangle) +
\cos^2\!\phi_j (\eta \langle\Delta\hat{p}^2_{s,j}\rangle +
                (1-\eta) \langle\Delta\hat{p}^2_{\mathrm{vac},j}\rangle)
\right) \\ &=
\frac{1}{M^2} \sum_{j=1}^M \left( \frac{\eta e^{2r_s}}{2} \sin^2\phi_j +
\frac{\eta e^{-2r_s}}{2} \cos^2\phi_j + \frac{1-\eta}{2} \right) .
\end{align}

The second equality comes from the fact that in the separable approach
there are no correlations between the modes. Under a stronger bound on
the magnitude of the phase shifts,
\(\phi_j \ll \sqrt{\langle\Delta\hat{p}^2_{s,j}\rangle / \langle\Delta\hat{x}^2_{s,j}\rangle}\),
this expression reduces to

\begin{equation}
\langle \Delta \hat{P}_\mathrm{avg}^2 \rangle \approx \frac{1}{M} \left( \eta\langle \Delta\hat{p}_{s,j}^2 \rangle + \frac{1-\eta}{2} \right)= \frac{\eta e^{-2r_s} + 1 - \eta}{2M} .
\end{equation}

Hence, the sensitivity is

\begin{equation}
\label{eq:sens_sep}
\sigma_s = \frac{\sqrt{e^{-2r_s} + 1/\eta - 1}}{2 \alpha_s \sqrt{M}} .
\end{equation}

The average number of photons hitting each sample is

\begin{equation}
N_s = N_{s,\mathrm{coh}} + N_{s,\mathrm{sqz}} = \eta(\alpha_s^2 + \sinh^2r_s) .
\end{equation}

\begin{figure}[htb]
\centering
\includegraphics[width=0.4\linewidth]{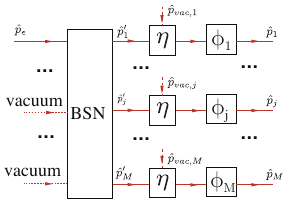}
\caption{Phase quadrature notations for analyzing the entangled scheme. The amplitude quadrature is defined accordingly. BSN: beam-splitter network with M inputs and outputs; $\hat p_e$: the only non-vacuum input of the BSN; $\hat p'_1\dots\hat p'_M$: the evenly split M output of the BSN. All the other notations are the same as Figure \ref{fig:sep}.}
\label{fig:ent}
\end{figure}

\subsubsection{Entangled scheme}\label{header-n34}

With entangled probes (the notation used in our analysis is summarized in Figure \ref{fig:ent}), we use the same estimator,
\(\hat{P}_\mathrm{avg}\). The individual modes that combine to form the
average are, however, now related through the distributed single initial
resource \(\hat{p}_e\):

\begin{equation}
\hat{p}_j = \left( \sqrt{\eta}\,\hat{x}'_j + \sqrt{1-\eta}\,\hat{x}_{\mathrm{vac},j} \right) \sin\phi_j  +  \left( \sqrt{\eta}\,\hat{p}'_j + \sqrt{1-\eta}\,\hat{p}_{\mathrm{vac},j} \right) \cos\phi_j,
\end{equation}
where the primed mode operators are obtained after symmetric
distribution in the beam-splitter network, that is,

\begin{equation}
\hat{x}_e = \frac{1}{\sqrt{M}}\sum_{j=1}^M \hat{x}'_j\ , \quad 
\hat{p}_e = \frac{1}{\sqrt{M}}\sum_{j=1}^M \hat{p}'_j.
\end{equation}

The mean value of the estimator is

\begin{equation}
\langle \hat{P}_\mathrm{avg} \rangle = \frac{1}{M} \sum_{j=1}^M \langle \hat{p}_j \rangle \approx \frac{\sqrt{\eta}}{M} \sum_{j=1}^M \langle \hat{x}'_j \rangle \phi_j = \sqrt{\frac{2\eta}{M}} \alpha_e \phi_\mathrm{avg} .
\end{equation}

The variance is

\begin{align}
\label{var_pavg_ent}
\langle \Delta \hat{P}_\mathrm{avg}^2 \rangle &= 
\frac{1}{M^2} \left\langle \Delta \left(\sum_{j=1}^M \hat{p}_j \right)^2 \right\rangle \nonumber\\
&\approx \frac{\eta}{M^2} \left\langle \Delta \left(\sum_{j=1}^M \hat{x}'_j \phi_j \right)^2 \right\rangle  +  
\frac{\eta}{M^2} \left\langle \Delta \left(\sum_{j=1}^M \hat{p}'_j \right)^2 \right\rangle +
\frac{1-\eta}{M^2} \left\langle \Delta \left(\sum_{j=1}^M \hat{p}_{\mathrm{vac}, j} \right)^2 \right\rangle
\nonumber\\
&\approx \frac{\eta}{M^2} \left\langle \Delta \left(\sum_{j=1}^M \hat{p}'_j \right)^2 \right\rangle  +  \frac{1-\eta}{M^2} \sum_{j=1}^M \langle \hat{p}_{\mathrm{vac},j}^2 \rangle \nonumber\\
&= \frac{\eta}{M^2} M \langle \Delta \hat{p}_e^2 \rangle + \frac{1-\eta}{M^2} \sum_{j=1}^M \langle \hat{p}_{\mathrm{vac},j}^2 \rangle \nonumber\\
&= \frac{\eta e^{-2r_e} + 1-\eta}{2M} .
\end{align}
In the second line, we made use of the fact that there are no correlations between $\hat{x}$ and $\hat{p}$ quadratures for the given probe state in our entangled scheme as well as the small angle approximation $\cos(\phi_j) \approx 1$, $\sin(\phi_j) \approx \phi_j$.
In the third line, we further tightened the small angle approximation by taking a $\tilde\phi$ such that for all $j$, $|\phi_j| < \tilde\phi$ and assuming
\begin{align}
    \tilde\phi^2 \left\langle \Delta \left(\sum_{j=1}^M \hat{x}'_j \right)^2 \right\rangle &\ll \left\langle \Delta \left(\sum_{j=1}^M \hat{p}'_j \right)^2 \right\rangle  \nonumber \\
    \Rightarrow \quad \tilde\phi^2 &\ll \frac{\langle \Delta \hat{p}_e^2 \rangle}{\langle \Delta \hat{x}_e^2 \rangle} \nonumber \\
    \Rightarrow \quad \tilde\phi &\ll e^{-2r_e} .
\end{align}
This approximation gives a sensitivity for the entangled approach of

\begin{equation}
\label{eq:sens_ent}
\sigma_e = \frac{\sqrt{\langle \Delta \hat{P}_\mathrm{avg}^2 \rangle}}
                    {\left| \partial \langle \hat{P}_\mathrm{avg} \rangle / \partial \phi_\mathrm{avg} \right|}
= \frac{\sqrt{e^{-2 r_e} + 1/\eta -1}}{2 \alpha_e} .
\end{equation}

Note that this, in contrast with the separable approach, does not depend
on the number of modes \(M\). The sensitivity is therefore the same as
the sensitivity for a single mode with the same resource state - but in
the single mode case the sample would of course be exposed to \(M\)
times as many photons. The average number of photons hitting each sample in the
distributed, entangled scheme is

\begin{equation}
N_e = N_{e,\mathrm{coh}} + N_{e,\mathrm{sqz}} = \frac{\eta}{M}(\alpha_e^2 + \sinh^2 r_e) .
\end{equation}

\subsection{Optimized parameters and sensitivities}\label{header-n48}

\subsubsection{Entangled scheme}\label{header-n49}

With the sensitivities given by eqs. \eqref{eq:sens_sep} and
\eqref{eq:sens_ent}, we wish to find the values for the displacement
amplitudes and squeezing parameters that optimize the sensitivity for a
fixed photon number on the sample. This problem can be solved with
Lagrangian multipliers, using the constraint \(N_{s,e} - N = 0\), where
\(N\) is the photon number to be held fixed during optimization. The
total photon number of the resource state(s) before loss is then
\(N_\mathrm{tot} = MN/\eta\).

The Lagrange function for the entangled scheme is

\begin{align}
\mathcal{L}_e(\alpha_e, r_e, \lambda) &= \sigma_e + \lambda (N_e - N) \\
&= \frac{\sqrt{e^{-2r_e} + 1/\eta - 1}}{2 \alpha_e} + \lambda \frac{\eta}{M}(\alpha_e^2 + \sinh^2r_e) - \lambda N ,
\end{align}
and the equations for the stationary point of the Lagrangian become

\begin{align}
0=\nabla_{\alpha_e}\mathcal{L}_e &= -\frac{\sqrt{e^{-2r} + 1/\eta - 1}}{2\alpha_e^2} + \frac{2\lambda \eta \alpha_e}{M}, \\
0=\nabla_{r_e}\mathcal{L}_e &= -\frac{e^{-2r_e}}{2\alpha_e \sqrt{e^{-2r_e} + 1/\eta - 1}} + \frac{2\lambda\eta \cosh r \sinh r_e}{M}, \\
0=\nabla_\lambda\mathcal{L}_e &= \frac{\eta}{M}(\alpha_e^2 + \sinh^2r_e)-N.
\end{align}

After some manipulation, the solutions can be expressed as

\begin{gather}
e^{2r_e} = \frac{\Lambda_M - \eta}{1-\eta}, \\
\alpha_e^2 = N_\mathrm{tot} - \sinh^2r_e = N_\mathrm{tot} - \frac{e^{2r_e} + e^{-2r_e} - 2}{4} 
= \frac{MN}{\eta} - \frac{(\Lambda_M-1)^2}{4(1-\eta)(\Lambda_M-\eta)},
\end{gather}
with \(\Lambda_M=\sqrt{1+4MN(1-\eta)}\). The optimal photon number ratio
is

\begin{equation}
\mu_e = \frac{N_{e,\mathrm{sqz}}}{N} = \frac{\sinh^2 r_e}{N_\mathrm{tot}} = \frac{\eta(\Lambda_M-1)^2}{4MN(1-\eta)(\Lambda_M-\eta)} ,
\end{equation}
and the optimal sensitivity obtained with these parameters becomes

\begin{equation}
\sigma_e^\mathrm{opt} = \frac{1}{2MN}\sqrt{\frac{MN(1-\eta) + \eta(\Lambda_M + 1)/2}{1 + \eta/(MN)}} ,
\end{equation}
which for \(\eta=1\) reduces to
\(\sigma_e^\mathrm{opt}(\eta=1) = \frac{1}{2MN} \sqrt{\frac{MN}{MN+1}}\).
This sensitivity exhibits Heisenberg scaling in both photon number (due
to the squeezing) and mode number (due to the entanglement).

\begin{figure}[ht]
\centering
\includegraphics[width=\linewidth]{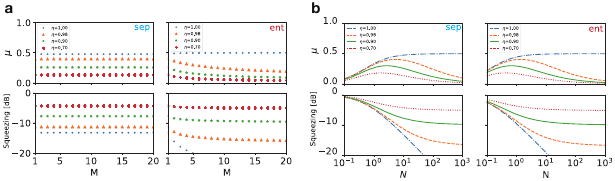}
\caption{
\textbf{a} and \textbf{b}: Optimized squeezed photon number ratio $\mu$ and squeezing degrees $\eta e^{-2r} + (1-\eta)$ for the Fig. 1\textbf{c} and 1\textbf{d} in the main text, respectively. Sep, the separable approach; Ent, the entangled approach.
}
\label{mu_sq}
\end{figure}

\subsubsection{Separable scheme}\label{header-n64}

Doing the same derivation for the separable scheme, that is, starting
from the Lagrange function
\(\mathcal{L}_s(\alpha_s,r_s,\tilde\lambda) = \sigma_s + \tilde\lambda(N_s-N)\),
results in the following optimal parameters for squeezing and
displacement:

\begin{gather}e^{2r_s} = \frac{\Lambda_1 - \eta}{1-\eta}, \\
\alpha_s^2 = \frac{N_\mathrm{tot}}{M} - \sinh^2r_s = \frac{N}{\eta} - \frac{(\Lambda_1-1)^2}{4(1-\eta)(\Lambda_1-\eta)},\end{gather}
with \(\Lambda_1=\sqrt{1+4N(1-\eta)}\), and a corresponding photon number ratio

\begin{equation}
\mu_s = \frac{N_{s,\mathrm{sqz}}}{N} = \frac{M\sinh^2 r_s}{N_\mathrm{tot}} = \frac{\eta(\Lambda_1-1)^2}{4N(1-\eta)(\Lambda_1-\eta)} .
\end{equation}

Finally, the optimal sensitivity becomes

\begin{equation}
\sigma_s^\mathrm{opt} = \frac{1}{2\sqrt{M}N} \sqrt{\frac{N(1-\eta) + \eta(\Lambda_1 + 1)/2}{1 + \eta/N}} ,
\end{equation}
which for \(\eta=1\) reduces to \(\sigma_s^\mathrm{opt}(\eta=1) = \frac{1}{2\sqrt{M}N} \sqrt{\frac{N}{N+1}}\), thus no longer showing Heisenberg scaling in the mode number. The result enable us to obtain the simulation result in Fig 1c and 1d in the main text, and the optimal $\mu$ and corresponding squeezing rated need to get the optimal $\mu$ is shown in Fig. \ref{mu_sq}.

%===================================
\section{Quantum Cram\'{e}r-Rao Bound for distributed phase sensing}
%===================================

In this section, we discuss the Quantum Cram\'{e}r-Rao Bound (QCRB) for $\phi_{avg}$ sensing, which defines the ultimate sensitivity limit for a given probe with multi-mode Gaussian state. We also compare the QCRB for $\phi_{avg}$ sensing to the counterpart of single parameter phase sensing by using a single-mode Gaussian state as probe, which was discussed in detail in Ref. \cite{Pinel2013PRA}.  Throughout the analysis, we assume the sensing channel has a constant efficiency $\eta$.

\begin{figure}[ht]
\centering
\includegraphics[width=0.6\linewidth]{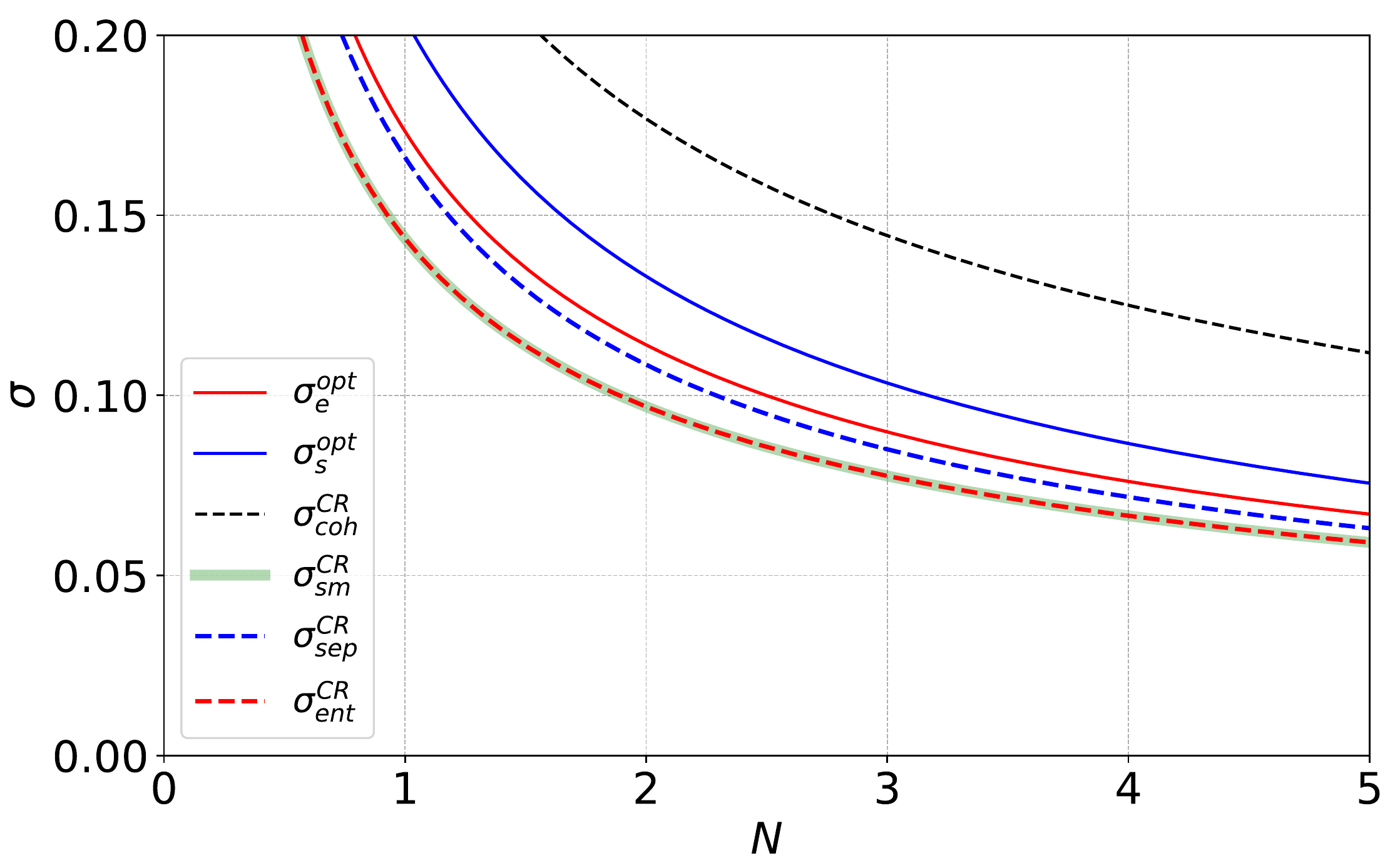}
\caption{
The optimal sensitivity of our separable scheme $\sigma_{s}^{opt}$ (solid blue) and entangled scheme $\sigma_{e}^{opt}$ (solid red) compared with the QCRB for different scenarios under our total efficiency $\eta$ = 0.735: The optimal QCRB for $\phi_{avg}$ sensing with coherent probes ($\sigma^{CR}_{coh}$, dashed black), the separable scheme with squeezed probes ($\sigma^{CR}_s$, dashed blue), and the entangled scheme ($\sigma^{CR}_e$, dashed red), as well as the QCRB for single parameter phase sensing with a squeezed probe ($\sigma^{CR}_{sm}$, solid green). Note that single parameter phase sensing is a quite different sensing task from $\phi_{avg}$ sensing, but $\sigma^{CR}_{sm}$ and $\sigma^{CR}_e$ coincides. 
}
\label{QCRB_plot}
\end{figure}

For a general sensing problem, the Quantum Cram{\'e}r-Rao Bound sets a lower bound, minimized over all possible measurements, on the uncertainty with which a parameter $\phi$ can be estimated through an unbiased estimator $\tilde{\phi}$, given a probe in a certain quantum state: $\langle \Delta \tilde\phi^2\rangle \geq F_\phi^{-1}$, where $F_\phi$ is the quantum Fisher information. The ultimate sensitivity limit for sensing of a single phase shift is thus $\sigma_\phi^{CR} = 1/\sqrt{F_\phi}$.
The quantum Fisher information for single mode phase sensing using a Gaussian probe with initial displacement $\alpha$ and squeezing $r$ is given by \cite{Pinel2013PRA}

\begin{equation}
    F_{sm}=\frac{e^{2r'}}{2N_{th}+1}\alpha^2 + \frac{(2N_{th}+1)^2 (e^{4r'}+e^{-4r'}-2)}{(4N_{th}+2)^2+4}
    \label{sm_qfi}
\end{equation}
with an effective thermalization photon number due to loss
\begin{equation*}
    N_{th} = \frac{1}{2} \left[(\eta e^{-2r}+1-\eta)(\eta e^{2r}+1-\eta)-1\right]
\end{equation*}
and and effective squeezing parameter
\begin{equation*}
    r' = \frac{1}{4} \log\left[\frac{\eta e^{2r}+1-\eta}{\eta e^{-2r}+1-\eta}\right] .
\end{equation*}
The QCRB for sensing of an average $\phi_\mathrm{avg}$ of multiple phase shifts with coherent probes, $\sigma^{CR}_\mathrm{coh}$, and with our separable approach, $\sigma^{CR}_\mathrm{sep}$, can also be found from Eq.~(\ref{sm_qfi}) divided by $\sqrt{M}$ to account for the $M$ independent phase estimations.

For non-trivial estimation involving multiple parameters, the quantum Fisher information matrix (QFIM) is needed.
The variance of an unbiased estimator $\tilde{q}$ of an arbitrary linear combination of 
$M$ parameters, $q = \sum_{i=1}^M w_i \phi_i$, is $\langle \Delta \tilde{q}^2\rangle = \mathbf{w}^T \mathbf{K} \mathbf{w}$ with the weight coefficients $\mathbf{w}^T = (w_1, \ldots, w_M)$ and parameter covariance matrix $\mathbf{K}$ with $K_{ij} = \langle (\tilde{\phi}_i - \phi_i)(\tilde\phi_k - \phi_j)\rangle$. Given a quantum Fisher information matrix $\mathbf{F}$, the QCRB is expressed as
\begin{equation}
    \langle \Delta \tilde{q}^2\rangle = \mathbf{w}^T \mathbf{K} \mathbf{w} \geq
    \mathbf{w}^T \mathbf{F}^{-1} \mathbf{w} = \sigma^{CR}_q.
\end{equation}
The QCRB for $\phi_\mathrm{avg}$ with our entangled scheme where the weights are $\mathbf{w}_\mathrm{avg}^T = (1/4,1/4,1/4,1/4)$ is then
\begin{equation}
    \sigma^{CR}_\mathrm{ent} = \mathbf{w_\mathrm{avg}^T} \mathbf{F}_\mathrm{ent}^{-1} \mathbf{w_\mathrm{avg}} 
    = \frac{1}{16}\sum_{i,j=1}^{4} (\mathbf{F}_\mathrm{ent}^{-1})_{ij}.  
\end{equation}
How to calculate the QFIM for arbitrary multi-mode Gaussian states, as well as the existence (or not) of a measurement that reaches the bound, is discussed in \cite{DominikJPA2018, NicholsPRA2018}.
We use Eqs. (16)-(21) in Ref. \cite{NicholsPRA2018} to numerically calculate $\mathbf{F}_\mathrm{ent}$.
As discussed in Ref. \cite{DominikJPA2018}, when any of the symplectic eigenvalues of the quadrature covariance matrix of the Gaussian state has unity value, the process in \cite{NicholsPRA2018} gives a singular result. This applies to our entangled scheme since it has 3 vacuum input modes. We solve this numerical problem pragmatically by replacing the 3 vacuum states with very weak thermal states (10$^{-6}$ mean photon number in each).

The optimal QCRBs for different scenarios together with the optimal sensitivity of our measurement schemes $\sigma^{opt}_s$ and $\sigma^{opt}_e$ are shown in Fig.~\ref{QCRB_plot}. The QCRBs are optimized over $\alpha$ and $r$ for a fixed mean photon number. It is interesting to note, that for the $\eta=0.735$ detection efficiency, the states optimizing the QCRBs are all squeezed vacuum states. From the result in Fig.~\ref{QCRB_plot}, we find at our detection efficiency: 
\begin{equation}
    \sigma^{CR}_{coh} > \sigma^{opt}_s > \sigma^{opt}_e > \sigma^{CR}_{sep} > \sigma^{CR}_{ent} .
\end{equation}
The sensitivities we obtain do not reach the corresponding ultimate limits. Furthermore, these relations show that in principle it should be possible to reach a better sensitivity with a separable scheme than what we obtain with the entangled scheme.
However, the difference between them is small and---to the best of our knowledge---no efficient way of experimentally implementing a measurement to reach the ultimate limit $\sigma^{CR}_{sep}$ is known. In \cite{Oh2019}, it is discussed that an $\hat{X}\hat{P}+\hat{P}\hat{X}$ type of measurement is needed. This is non-Gaussian and can not be realized by only Gaussian operations such as squeezing, beam splitters, phase shifts and homodyne/heterodyne detection.
For sensing involving multiple parameters, it is unclear how to reach the optimal bound $\sigma^{CR}_\mathrm{ent}$---or whether it is at all possible. A joint, non-Gaussian measurement may be optimal. 

Finally, note that the QCRB of the entangled scheme overlaps with that of the single parameter estimation. This makes sense intuitively: Splitting a resource state equally in four and using these to probe the average of four phase shifts should result in the same sensitivity as that of a single phase shift probed by the same un-split resource.

\section{Preparation of entangled probes}

The entangled probes are prepared in two steps. First, we generate a squeezed coherent state, denoted as the squeezed probe (SP), by an optical parametric oscillator (OPO). Second, we send the SP through a beam-splitter network (BSN) to generate 4 entangled probes. 
We define the mode of the SP to be a narrow sideband at 3 MHz, since this is region where we have high squeezing quality.

\subsection{Generation of squeezed probes with OPO}

\begin{figure}[ht]
\centering
\includegraphics[width=0.8\linewidth]{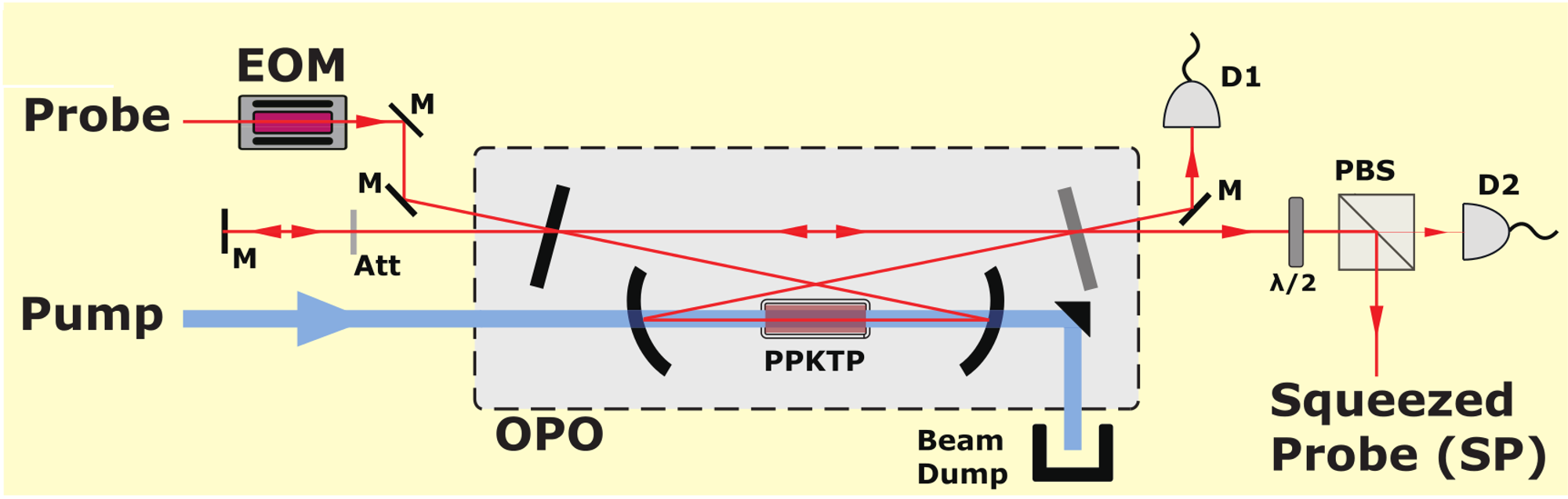}
\caption{Squeezed probe (SP) preparation with OPO. M, high reflectively mirror; EOM, electro-optic modulator; Att, attenuator; D1, resonant detector for cavity lock; D2, high gain detector for OPO gain lock.}
\label{fig:OPO}
\end{figure}

The laser source for the experiment is an amplified NKT Photonics X-15 fibre laser operating at 1550 nm. Most of the light is used for pumping a second harmonic generation (SHG) cavity (same design as the OPO described below) to produce 775 nm light to act as the OPO pump. The rest is used for the local oscillator and the probe and lock beams.
As shown in Fig. \ref{fig:OPO}, we use a bow-tie shaped OPO with a periodically poled potassium titanyl phosphate (PPKTP) crystal to generate the SP by type-0 parametric down conversion. The bandwidth of the cavity is 8.0 MHz half width half maximum (HWHM) and the OPO pump power threshold is 850 mW. The 775 nm pump, which for the measurements presented here varied between 150 mW and 350 mW, is coupled through the dichroic curved cavity mirrors and dumped after passing the crystal. A 3.6 mW coherent beam at 1550 nm, weakly phase-modulated by an electro-optic modulator (EOM) at 3 MHz and 28.7 MHz, is coupled into the OPO in the counter-propagating direction through a high reflectivity mirror (HR) with a transmittance of about 100 ppm. This beam is used to lock the cavity by the Pound--Drever--Hall technique with the 28.7 MHz side band and the resonant detector D1.
All cavity and phase locks in the experiment are handled by Red Pitaya FPGA boards running the PyRPL lockbox software \cite{pyrpl}.

The reflection from the HR mirror is re-coupled into the forward-propagating mode of the OPO with a 0$^\circ$ mirror to serve as the carrier of the sideband mode that defines our probe state. A variable attenuator (Att.) is inserted to control the optical power. In the OPO, the forward-propagating beam is squeezed by the parametric process and coupled out through a 10$\%$ transmittive out coupling mirror (light gray in Fig. \ref{fig:OPO}). A half-wave plate ($\lambda/2$) and a polarization beam-splitter (PBS) is used to tap around $1\%$ of the OPO output towards detector D2 to lock the phase between the carrier and the pump for de-amplification. As a result, the carrier is squeezed in the amplitude quadrature, leading to squeezing of the phase quadrature of the probe in the 3 MHz modulated sideband frequency mode since the sideband is encoded by phase modulation.

\subsection{Generation and detection of entangled probes}

\begin{figure}[ht]
\centering
\includegraphics[width=0.8\linewidth]{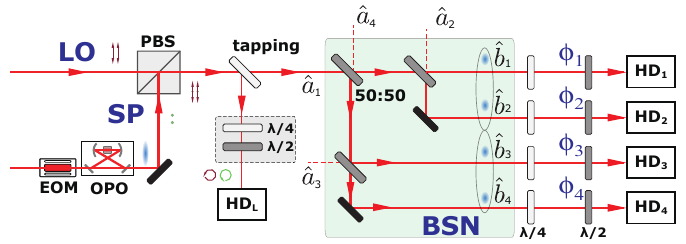}
\caption{Detailed experimental setup and the input-output relationship of the beam splitter network (BSN).
The input modes $\hat a_1$ to $\hat a_4$ with s-polarization are transferred into the output modes $\hat b_1$ to $\hat b_4$. Here, only $\hat a_1$ is a squeezed coherent state operator. $\hat a_2$ to $\hat a_4$ are vacuum operators. By tuning the wave plates at each output of the BSN, $\hat b_1$ to $\hat b_4$ are set to circular polarization.}
\label{fig:BSN}
\end{figure}

The detailed experimental setup is shown in Fig. \ref{fig:BSN}. It is essentially a multi-port version of a squeezed-light-enhanced polarization interferometer \cite{Grangier1987}. We create four entangled probes by sending the squeezed probe, SP, through a BSN consisting of three 50:50 beam-splitters.  
Prior to this, the SP is spatially combined on a PBS with a strong beam (LO) which will act as the local oscillator for all four modes.
%The SP and the local oscillator (LO) with orthogonal linear polarizations are spatially combined on a PBS. 
The LO phase is locked to either the $\hat{p}$ or $\hat{x}$ quadrature of the SP by tapping $\sim 1\%$ towards a polarization-based homodyne detection setup, the output of which is used to control a piezo-mounted mirror in the LO path.
In each of the four modes, the phase between LO and SP can be further controlled by a $\lambda/4$ and a $\lambda/2$ wave plate.
The $\lambda/4$ plates change the LO and SP into left-hand and right-hand circular polarization, respectively.
The $\lambda/2$ plates introduce phase shifts between SP and LO and play two roles: 
First, they are used to synchronize the phases for the entangled probes by compensating the phase difference induced by 50:50 beam splitters.
Second, they are used to simulate the phase samples, that is, the imposed phases $\phi_1, \dots, \phi_4$.
For details, see section \ref{sec:phasecontrol}.
Finally, the four outputs are measured on homodyne detectors.

\subsubsection{\label{sec:HD}Homodyne detection and data acquisition}

\begin{figure}[ht]
\centering
\includegraphics[width=0.98\linewidth]{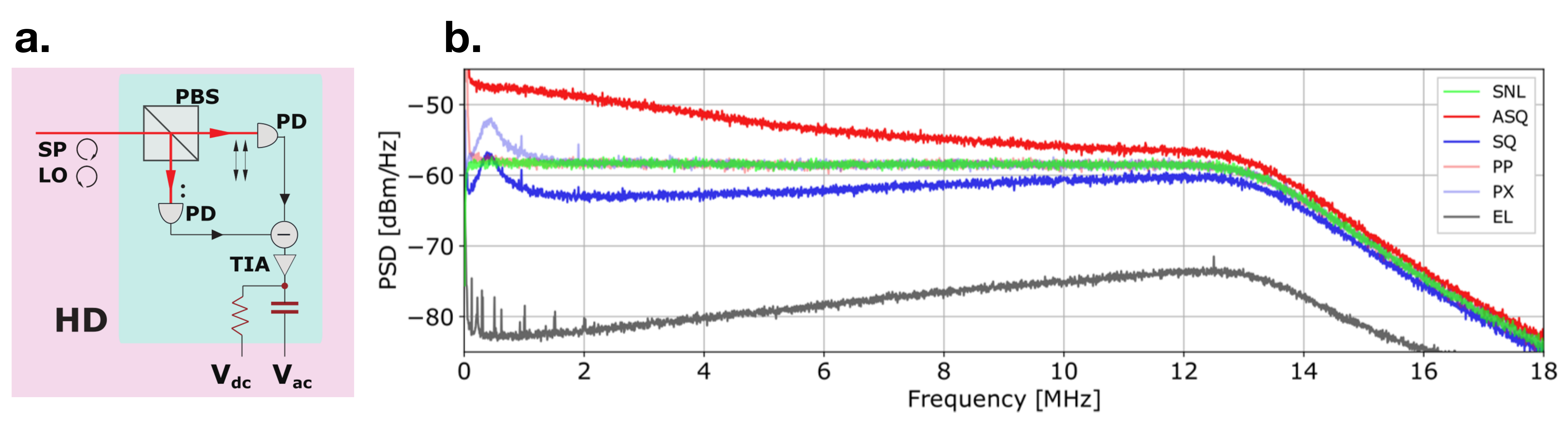}
\caption{\textbf{a.} The circular polarized SP and local oscillator (LO) are projected into p- or s-polarization by a PBS, and detected with a balanced photo-detector. The output voltage of the detector photo diodes is separated into DC--100 kHz output $V_{dc}$ and 100 kHz--14 MHz output $V_{ac}$.  PD, photo diode; TIA, trans-impedance amplifier. \textbf{b.} Power spectral densities (PSDs). SNL, shot-noise limit; ASQ, anti-squeezing; SQ, squeezing; PX and PP, probe noise in X and P quadrature respectively, measured by blocking pump; EL, electronic noise of the data acquisition system, measured by blocking both SP and the LO.}
\label{fig:HD}
\end{figure}

All five homodyne detection setups use the same scheme, illustrated in Fig. \ref{fig:HD}a. The circularly polarized SP and LO interfere after the PBS. The optical power of the LO is about 3 mW on each HD and it detects a SP of about 10 nW. The output of the detector is electronically split into AC and DC parts with a bias-tee of about 100 kHz. The AC signal, $V_{ac}$, is used for phase sensing. It includes the 3 MHz side-band, but filters out the carrier at DC and the side-band for cavity locking at 28.7 MHz with a low pass filter at around 14 MHz. The DC signal, $V_{dc}$, detects the carrier. It is used for phase locking and phase calibration (see section \ref{sec:phasecalibration}).

The $V_{ac}$ outputs of HD1 to HD4 are sent to a 4-channel oscilloscope (LeCroy HDO6034), which acquires time-voltage traces of 200 $\mu$s with a 50 MHz sampling rate. 
The power spectral densities (PSDs) of the individual HD outputs and the averaged output is obtained by Fast Fourier Transform (FFT) on a computer.
Fig. \ref{fig:HD}b shows PSDs for the averaged voltage of the 4 HDs in different experimental conditions with no modulation from the EOM. All the PSDs in Fig. \ref{fig:HD}b are the averaged result of 400 oscilloscope measurements.  
To show the signal-to-noise ratio of the data acquisition system, we measure the PSDs of the shot noise level (SNL, measured when SP is blocked) and electronic noise (EL, measured when both SP and LO are blocked). The result is shown in green and dark grey traces in Fig. \ref{fig:HD}b. 
We see that the electronic noise clearance is about 23 dB at the 3 MHz side band, which corresponds to about $0.5\%$ effective loss in detection efficiency. 
We will discuss the other PSDs shown in Fig. \ref{fig:HD}b in the following subsection.

\subsubsection{Input-output relations of the BSN}

The BSN we use in the experiment is shown in Figure \ref{fig:BSN}.  The only non-vacuum input mode $\hat a_1$ is the SP, whose mode operator is $\hat a_1=\hat S^\dagger(r)  \hat a \hat S(r) + \alpha$ in the Heisenberg picture, with $\hat a$ being the annihilation operator of the OPO input at 3 MHz and the real-valued $\alpha$ being the effective coherent excitation of the mode after modulation by the EOM and de-amplification in the OPO. 
All the other input modes $\hat a_2$, $\hat a_3$ and $\hat a_4$ are vacuum modes. 
The output modes of the BSN, $\hat{b}_j$ can be explicitly written as:

\begin{eqnarray}
\label{eq:modes}
\hat b_{1} &=& \frac{1}{2} \sqrt{\eta} (\hat a_1 -i\hat a_4+\sqrt{2}i\hat a_2)+\sqrt{1-\eta}\hat a_{vac,1} \nonumber \\
\hat b_{2} &=& \frac{1}{2} \sqrt{\eta} (\hat a_1 -i\hat a_4-\sqrt{2}i\hat a_2)+\sqrt{1-\eta}\hat a_{vac,2} \nonumber \\
\hat b_{3} &=& \frac{1}{2} \sqrt{\eta} (\hat a_1 +i\hat a_4+\sqrt{2}i\hat a_3)+\sqrt{1-\eta}\hat a_{vac,3} \nonumber \\
\hat b_{4} &=& \frac{1}{2} \sqrt{\eta} (\hat a_1 +i\hat a_4-\sqrt{2}i\hat a_3)+\sqrt{1-\eta}\hat a_{vac,4}.
\end{eqnarray}
Here we have introduced an identical overall efficiency $\eta$ and vacuum mode operator $\hat a_{vac,j}$ for $j=1\dots4$. 
Although the various inefficiencies occur at different points in the experiment, for simplicity we have assumed (as in section \ref{sec:sensitivity}) that they all occur after the distribution of the probes in the BSN and that they are identical for the four channels. Experimentally, we use eight variable irises before the PDs of all four HDs to equalize the overall detection efficiency.

\subsubsection{Overall detection efficiency estimation}
The loss budget of our experiment setup is as follows: the escape efficiency of the OPO $\sim95\%$; the quantum efficiency of the photo diodes in HD $\sim98\%$; the imperfection of the mode matching between SP and LO  $\sim90\%$; the electronic noise of the homodyne detection $\sim99\%$; the efficiency introduced by tapping for phase locking $\sim97\%$ and the efficiency of all optics between OPO output and the PD of the HD $\sim 92\%$. The loss budget of the experiment system gives an estimation of the overall detection efficiency of $\eta\sim74\%$.

We also estimate the overall detection efficiency by measuring the squeezing/anti-squeezing degrees (notated with $v^2_{sq}$ and $v^2_{asq}$) for the entangled approach at 3 MHz. Since
\begin{eqnarray}
&v^2_{sq} = \eta e^{-2 r_e} + (1-\eta)\nonumber\\
&v^2_{asq}= \eta e^{2 r_e} + (1-\eta),
\end{eqnarray}
we can calculate $\eta$ and $r_e$ with measured $v^2_{sq}$ and $v^2_{asq}$. The overall efficiency estimated with 5 different pump powers to the OPO is $\eta=73.5\%\pm1.5\%$. This result coincide with the loss budget estimation, and we use this result to theoretically calculate the sensitivity. 

For the separable approach, where the BSN is removed, the overall efficiency is $\sim1.5\%$ higher. However, we compensate this by tapping more to the lock detector D2 in Figure \ref{fig:OPO} so the separable approach has similar efficiency to that of the entangled approach.

\subsubsection{Entanglement characterization of the probes}

The squeezing degree for each individual output mode will not be better than 3/4 shot noise due to the splitting of the SP in the BSN. However, the squeezing of SP is converted into entanglement between all the probes. By joint measurement of the 4 probes (simply averaging the voltage from the four HDs), we can recover the squeezing degree of the SP: 
From Eq. (\ref{eq:modes}), the joint measurement recovering the squeezing of SP is simply the sum of the four HD$_\text{1--4}$ outputs. 
The recovered squeezed and anti-squeezed quadratures are shown as SQ (blue) and ASQ (red) in Fig. \ref{fig:HD}b. We see the joint measurement gives about 4.8 dB of squeezing at the 3 MHz side band frequency. The additional noise seen below 2 MHz is due to technical noise from our laser.
As a calibration of the noise of the probe before the parametric process, we measure the PSDs of $\hat X$ and $\hat P$ quadrature by blocking the pump of our OPO, and the result is shown with PX (light blue) and PP (light red) in Fig. \ref{fig:HD}b (noting that here we refer to the amplitude/phase quadrature of the carrier of the SP since there is no side-band). We see the technical noise of both $\hat X$ and $\hat P$ quadrature decreases as the frequency increases and overlap with the SNL when the frequency is above 1.8 MHz. Therefore, in our estimation of the overall detection efficiency at the side band frequency (3 MHz), we ignore contributions from technical noise of the laser.

\begin{figure}[ht]
\centering
\includegraphics[width=0.8\linewidth]{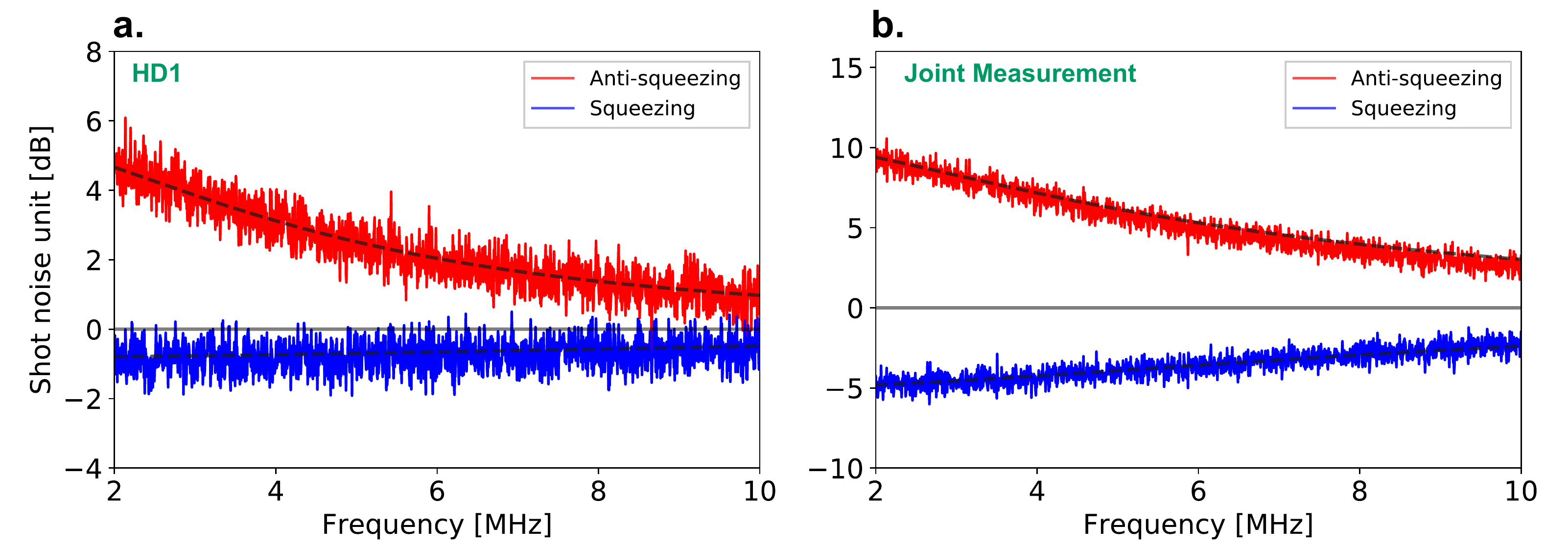}
\caption{
\textbf{a. } Squeezing and anti-squeezing spectra for a single distributed spatial mode obtained from HD$_1$. \textbf{b. } Squeezing and anti-squeezing spectra from joint measurement. Dashed lines: spectra predicted by theory. 
}
\label{fig:spec}
\end{figure}

With the measurement described in Fig. \ref{fig:HD}b, we can get the squeezing/anti-squeezing degree in SNL units. Fig. \ref{fig:spec}a and b shows the squeezing and anti-squeezing of an individual channel (HD1) and that from the joint measurement, respectively. The dashed lines show the squeezing and anti-squeezing predicted by \cite{polzik1992}
\begin{equation}
    S^{\pm}(f) = 1 \pm \frac{4 \eta \sqrt{{P}/{P_{th}}} }{(1 \mp \sqrt{{P}/{P_{th}}})^2 + (f/f_{cav})^2}, 
\end{equation}
where $S^{-}(f)$ and $S^{+}(f)$ denotes the squeezing and anti-squeezing  spectrum, $\eta = 0.735$ is the estimated overall detection efficiency, $f_{cav} = 8.0$ MHz is the HWHM of the OPO cavity and $P_{th}=850$ mW is the threshold of the OPO. In this measurement, $P = 300$ mW pump power is used. Here both $f_{cav}$ and $P_{th}$ are obtained from independent measurements.

We quantitatively verify the entanglement of the probes by reconstructing the covariance matrix of the 4 modes.
As we do not expect correlations between $\hat{x}$ and $\hat{p}$ quadratures, we only experimentally reconstruct $M_x=\mathrm{Cov}(\hat x_j, \hat x_k )$ and $M_p=\mathrm{Cov}(\hat p_j, \hat p_k )$ for $j,k=1$ to $4$ at around 3 MHz. 
After balancing the length of cables from HD$_\text{1--4}$ to the oscilloscope, we digitally filter the recorded traces by a 50 kHz band pass filter centered around 3 MHz, and measure $M_x$ and $M_p$, respectively. The covariance matrices in shot noise units from the average of 400 oscilloscope measurements are:
\begin{equation}
\label{cov}
	{M_{x}}=\begin{pmatrix}
		0.83 & -0.18 & -0.17 & -0.19\\
		\text{-} & 0.84 & -0.16 & -0.18\\
		\text{-} & \text{-} & 0.83 & -0.18\\
		\text{-} & \text{-} & \text{-} & 0.82
	\end{pmatrix}  \quad
	{M_{p}}=\begin{pmatrix}
		3.0 & 1.9 & 1.9  & 2.0\\
		\text{-} & 2.8 & 1.8 & 1.9\\
		\text{-} & \text{-} & 2.8 & 1.9\\
		\text{-} & \text{-} & \text{-} & 3.0
	\end{pmatrix},
\end{equation}
where symmetric elements are not shown. We show the entanglement property of the probes by calculating the logarithmic negativity $\mathcal{N}(\hat{\rho})$ between them, where $\mathcal{N}(\hat{\rho})>0$ is a sufficient condition for entanglement \cite{Vidal2002}. 
For a Gaussian state this can be obtained through the symplectic eigenvalues of the partially transposed covariance matrix, so that $\mathcal{N}(\rho)=\sum_k f(\tilde{v}_k)$, where $\tilde{v}_k$ are the symplectic eigenvalues and $f(x)=-\mathrm{log}_2(x)$ for $x<1$ and 0 otherwise. By constructing the full $M_{x,p}$ covariance matrix from $M_x$ and $M_p$, we find that for any two, three or four modes the value of $\mathcal{N}(\hat{\rho})$ is within the range of $0.20\pm0.02$, $0.33\pm0.02$ and $0.51\pm0.02$ respectively, confirming the presence of quadrature entanglement across all mode combinations.

\section{\label{sec:phasecal}Phase control and calibration}

In this section we first calculate the interference at the two photo diodes of the HD in Fig. \ref{fig:HD}. The result shows that the phase between SP and LO can be controlled by rotating the $\lambda/2$ wave plates. After that, we describe the phase calibration procedure and result in our experiment. With the phase calibration result, we can control the phase $\phi_j$ for $j=1\dots4$ (and therefore $\phi_\mathrm{avg}$) by rotating the $\lambda/2$ wave plates to a specific position.

\subsection{\label{sec:phasecontrol}Phase control with $\lambda/2$ wave plates}
The LO with p polarization and OPO output with s polarization are combined by the PBS in Fig. \ref{fig:BSN}, and the Jones vector after the PBS is
\begin{equation}
J_{in} = \begin{bmatrix} E_{LO} \cdot e^{-i\phi_{LO}} \\ E_{SP} \cdot e^{-i\phi_{SP}}\end{bmatrix},
\end{equation}
where $E_{LO} \cdot e^{-i\phi_{LO}}$ is the LO and $E_{SP} \cdot e^{-i\phi_{SP}}$ is the OPO output (squeezed probe). The Jones Matrix for a wave plate is \cite{Collet2005_book}
\begin{equation}
\label{eq:waveplate}
M_{wp} = 
\begin{bmatrix} 
\cos(\phi/2)+i\sin(\phi/2)\cos(2\theta) &  i\sin(\phi/2)\sin(2\theta)\\
i\sin(\phi/2)\sin(2\theta)              &  \cos(\phi/2)-i\sin(\phi/2)\cos(2\theta)
\end{bmatrix},
\end{equation}
where $\theta$ is the angle between the fast axis of the wave plate and p polarization (the direction of LO), and $\phi$ is the retardance of the wave-plate ($\phi=\pi$ or $\phi=\pi/2$ for an ideal $\lambda/2$ or $\lambda/4$ wave-plate, respectively). We fix the $\lambda/4$ wave plate at $\theta=45^\circ$ and put the $\lambda/2$ wave plate at a variable angle $\theta_{v}$, resulting in the output Jones vector 
\begin{equation}
    J_{out}=M_{\lambda/2}(\theta_v)M_{\lambda/4}(45^\circ)J_{in}  = \begin{bmatrix} J_1 \\ J_2\end{bmatrix}
\end{equation}
with

\begin{eqnarray}
J_1 &=& \frac{1}{\sqrt{2}} \left[i E_{LO} e^{i( 2\theta_v - \phi_{LO})}  - E_{SP} e^{-i( 2\theta_v  + \phi_{SP})} \right] \nonumber\\
J_2 &=& \frac{1}{\sqrt{2}} \left[E_{LO} e^{i( 2\theta_v - \phi_{LO})}  - i E_{SP} e^{-i( 2\theta_v  + \phi_{SP})} \right].
\end{eqnarray}
Therefore, the interference between the two polarization modes observed at the two diodes of the HD after the second PBS is:
\begin{equation}
\label{eq:hd_intensity}
I_{HD} = |J_1|^2- |J_2|^2= 2E_{SP} E_{LO} \sin(4\theta_{v} - \phi_d ),    
\end{equation}
where $\phi_d=\phi_{SP}-\phi_{LO}$ is the initial phase difference between OPO output and LO mode before being overlapped at the first PBS. The result show that if we rotate the $\lambda/2$ wave plate by an angle of 1$^\circ$, the phase between LO and OPO output will change 4$^\circ$. The form of Eq. (\ref{eq:hd_intensity}) also shows the visibility of the HD is not affected by the polarization transformation since it doesn't have any constant term. However, if the wave plates or PBS are not perfect, which means that the wave plates have either more or less retardance or that the PBS has a finite extinction ratio between s and p polarization, a similar calculation shows the rotation of $\lambda/2$ wave plate by 1$^\circ$ will result in a phase shift slightly deviating from 4$^\circ$, and that the visibility of the interference at HD can be reduced. We experimentally measure these imperfections as shown in the following subsection.

\subsection{\label{sec:phasecalibration}Phase calibration}

\begin{figure}[ht]
\centering
\includegraphics[width=0.85\linewidth]{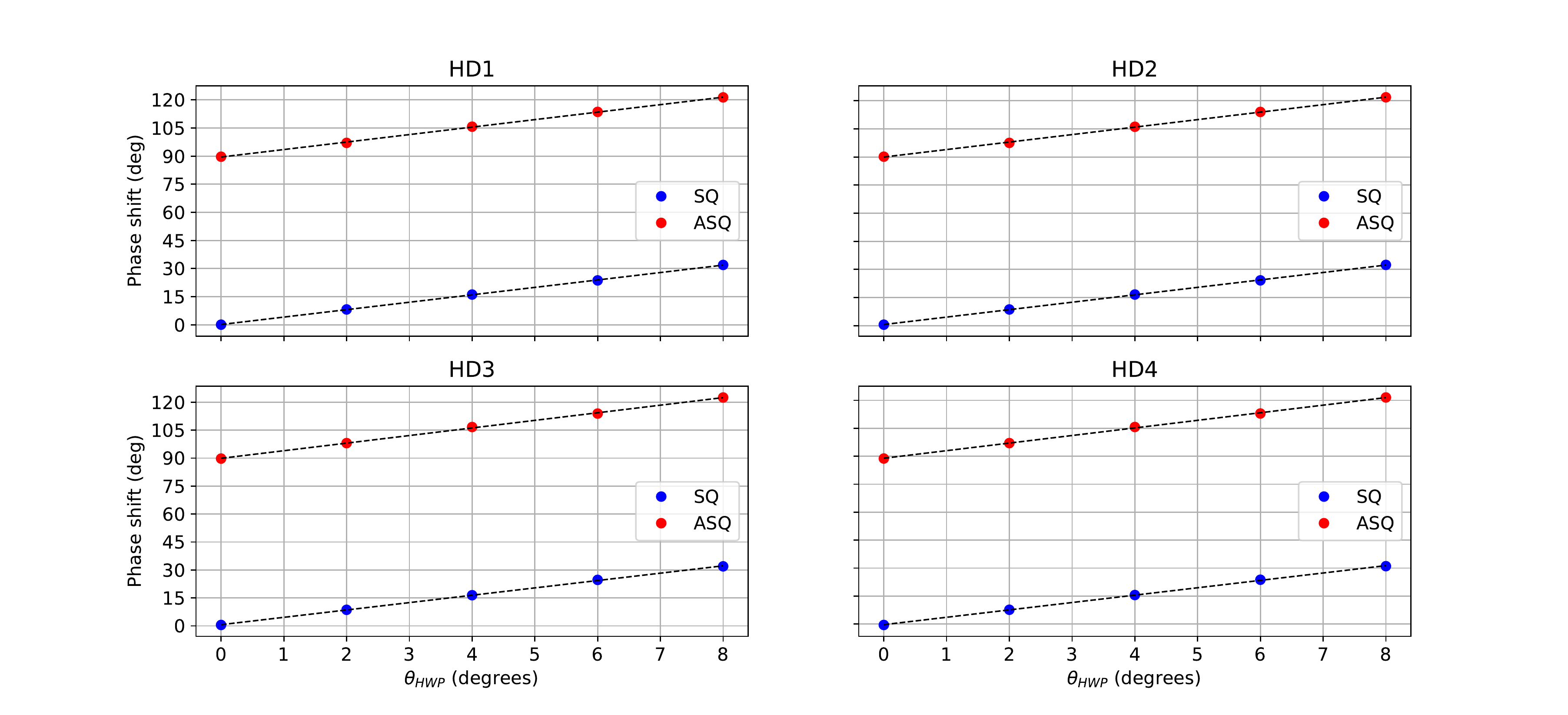}
\caption{Phase calibration result for each HD. The black dashed line is the linear fitting of the calibration. }
\label{fig:phasecal_result}
\end{figure}

During the experiment we lock $ \phi_d$ to be either $0^\circ$ or $90^\circ$ with HD$_L$, and use the rotation of the $\lambda/2$ wave plate before each HD to control the phase of each mode. In order to account for potential imperfections in our experiment, we first measured the visibility reduction from imperfect polarization components. We find a worst-case reduction of the HD visibility from $98.5\%$ to $95.2\%$. We also perform a phase calibration by scanning the phase between LO and SP carrier with a ramp at 27 Hz while the interference fringe measured from $V_{dc}$ of HD$_L$ and HD$_{1,2,3,4}$ is recorded. The phase between LO and signal in each path is inferred from sine curve fitting. We calibrate the phase with 40 repeated measurements at each $\lambda/2$ wave plate position, and the result is shown in Fig. \ref{fig:phasecal_result}. The SQ (blue dots) shows the result when we lock $\phi_i =0^\circ$ and the HDs measure the squeezed quadrature, and the ASQ (red dots) shows the result when we lock $\phi_i =90^\circ$. For both SQ and ASQ, we rotate the $\lambda/2$ wave-plate position in each channel by an actuator in the wave-plate mount, allowing us to faithfully use the calibration result in the experiment. 

\begin{table}[h!]
\caption{\label{tab:phasecal}
The phase calibration result}
\begin{tabular}{|l|l|l|l|l|l|l|l|}
\hline
\multicolumn{8}{|l|}{Squeezing}                                            \\ \hline
k1 & -3.96  & k2 & -3.97  & k3 & -3.95  & \multicolumn{1}{r|}{k4} & -3.96  \\ \hline
b1 & -0.13  & b2 & -0.59  & b3 & -0.64  & b4                      & 0.37   \\ \hline
\multicolumn{8}{|l|}{Anti-squeezing}                                       \\ \hline
k1 & -3.99  & k2 & -3.99  & k3 & -4.06  & k4                      & -4.06  \\ \hline
b1 & -89.50 & b2 & -89.97 & b3 & -89.87 & b4                      & -88.90 \\ \hline
\end{tabular}
\end{table}

From the calibration results, we see that the phase is linear within the whole range of the actuator (8$^\circ$) on the wave plate mounts. The result of the linear fitting to HD channel $j=1$ to $4$ with the equation 
\begin{equation}
    \phi_j = k_j \theta_{v} + b_j
\end{equation}
is summarized in Table \ref{tab:phasecal}. With these fitted parameters, we can control both the phase in each channel $\phi_j$ or the averaged phase $\phi_\mathrm{avg}$ accurately. Particularly, if we lock $ \phi_i$ to $0^\circ$, we can change the $\phi_\mathrm{avg}$ by a slope of $3.96^{\circ}\pm 0.02^{\circ}$; if we lock $ \phi_i$ to $90^\circ$, can change the $\phi_\mathrm{avg}$ by a slope of $4.02^{\circ}\pm 0.02^{\circ}$.

\section{Data analysis}

In this section we introduce the details of our data analysis procedure, which includes measuring the sensitivity by fitting and counting how many photons in average is used in the SP.
%Fig. \ref{fig6} shows the PSD results measured in different $\phi_\mathrm{avg}$, where each PSD result for sensitivity estimation is from the average 2000 oscilloscope measurements.

\begin{figure}[ht]
\centering
\includegraphics[width=0.6\linewidth]{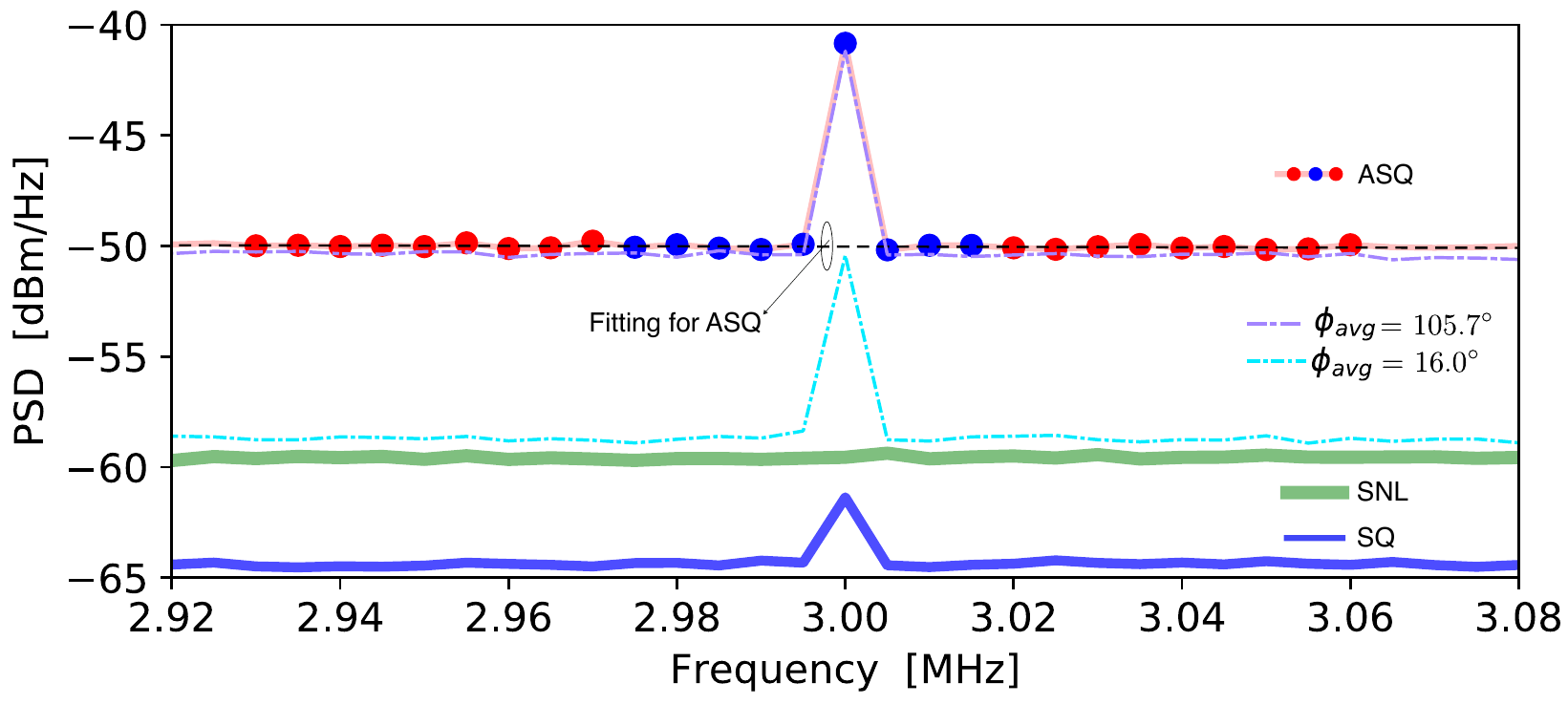}
\caption{ PSDs of averaged HD output voltage with 3 MHz phase modulation on at different $\phi_\mathrm{avg}$. SNL: shot-noise limit; SQ,  $\phi_\mathrm{avg}$=0.2 $\pm0.8^\circ$; ASQ, $\phi_\mathrm{avg}$ = 89.5 $\pm0.8^\circ$. We estimate $V_s$ and $V_n$ from these PSDs.
}
\label{fig6}
\end{figure}

As the estimator of $\phi_\mathrm{avg}$, $\hat P_\mathrm{avg}$ is experimentally estimated from the PSD of the averaged output of the four HDs in each mode. Fig. \ref{fig6} shows the PSD results measured for different $\phi_\mathrm{avg}$.
%where each PSD result for sensitivity estimation is from the average 2000 oscilloscope measurements. 
Each PSD is obtained from the FFT of an average of 2000 oscilloscope traces.
The spectrum peak at 3 MHz $S_{pk}$ gives the value of
\begin{equation}
 \label{spk}
 S_{pk}= 2 V^2_{sn} \cdot \langle \hat P_\mathrm{avg}^2 \rangle = 2 V^2_{sn} \cdot (\langle  \Delta \hat P_\mathrm{avg}^2 \rangle + \langle \hat P_\mathrm{avg} \rangle^2 ),
\end{equation}
where $V_{sn}$ is the 4-mode shot noise limit (SNL) voltage from HDs decided by LO power, electronic gain and the digital filtering. The constant 2 in Eq. (\ref{spk}) comes from the commutation relationship we choose $[\hat X, \hat P] = i$.
We start our data analysis by separating the peak into two voltage parts 
\begin{equation}
    S_{pk} = V^2_s + V^2_n,
\end{equation}
where $V_s = \sqrt{2} V_{sn} |\langle \hat P_\mathrm{avg}\rangle| $ is the signal part induced by the coherent photons of the side band, and $V_n = \sqrt{2} V_{sn} \sqrt{ \langle \Delta  \hat P_\mathrm{avg}^2 \rangle }$ is the part induced by the fluctuation of the light. 
Except at the 3 MHz peak, the spectra in Fig. \ref{fig6} vary slowly with frequency. This enables us to extract $V_n$ from the adjacent frequencies of the 3 MHz peak. The procedure of $V_{n}$ estimation is illustrated with the anti-squeezing quadrature (ASQ, $\phi_\mathrm{avg}$ = -89.5 $\pm0.8^\circ$) PSD in Fig. \ref{fig6} as an example. 
We first do a linear fit with the frequency range indicated by the red dots, which is slightly away from 3 MHz. This fitting gives the black dashed line labeled as "Fitting for ASQ". $V_n$ is then inferred by the square root value of the fitted line at 3 MHz. Since our side band line width is obviously smaller than the 5 kHz resolution of the FFT, only one peak point is observed in the PSDs in Fig. \ref{fig6}. Therefore, $V_s$ can simply be calculated by the difference between the blue dot at 3 MHz and the fitting result.

In our experiment we always introduce equal positive phase shift in all channels. In this case we know that $\langle \hat P_\mathrm{avg}\rangle>0$, and $V_s$ and $V_n$ relate to the averaged phase $\phi_\mathrm{avg}$ by 
\begin{eqnarray}
\label{eq:VsVn}
V_s (\phi_\mathrm{avg}) &=& \sqrt{2} V_{sn} \langle \hat P_\mathrm{avg}\rangle = \frac{2}{\sqrt{M}} V_{sn} \cdot \alpha_e |\sin (\phi_\mathrm{avg} + \theta_1)| \nonumber \\
V_n (\phi_\mathrm{avg}) &=& \sqrt{2} V_{sn} \sqrt{\Delta \hat P^2_\mathrm{avg}} = V_{sn}  \cdot  \sqrt{ v^2_{sq} \cos^2(\phi_\mathrm{avg} + \theta_2) + v^2_{asq} \sin^2(\phi_\mathrm{avg} + \theta_2) }.
\end{eqnarray}
Here $\alpha$ is the real coherent amplitude from modulation, M is the mode number, $\theta_{1}$ and $\theta_{2}$ are parameters indicating the imperfections of the experimental setup (ideally they should be 0), where $\theta_{1}$ parametrizes the residual amplitude modulation of the phase modulating EOM and $\theta_{2}$ parametrizes the phase locking offset of the squeezing measurement. $v^2_{sq}  =  \eta e^{-2r_e} + (1-\eta)$ and $v^2_{asq}  =  \eta e^{2r_e} + (1-\eta)$ are squeezing and anti-squeezing degrees in SNL units.
Note that the form of Eq. (\ref{eq:VsVn}) rely on two assumptions: First, we assume that the modulation signal on the EOM is perfectly coherent so $V_s(\phi_\mathrm{avg})$ doesn't have an offset term. This assumption is consolidated by the fact that we drive the EOM with a sine wave generated from a function generator with phase noise less than -65 dBc. Second, we ignore the phase fluctuations of the phase locking. This assumption is consolidated by the high ($\sim$32 dB) signal-to-noise ratio of the locking detector HD$_L$, though this signal-to-noise ratio is not a direct measurement of the phase fluctuation.

\subsection{Sensitivity fitting}

With $V_{s}$ and $V_{n}$ extracted for a range of $\phi_\mathrm{avg}$ settings, we can estimate the sensitivity. By comparing the definition of $\sigma$ in Eq. (\ref{eq:sens_def}) with Eq. (\ref{eq:VsVn}), the sensitivity to a small phase shift at a given $\phi_\mathrm{avg}$ offset is 
\begin{equation}
\label{sigma_vn_vs}
\sigma = V_n (\phi_\mathrm{avg}) / V'_s(\phi_\mathrm{avg}),
\end{equation}
where $V'_s = \partial V_s/\partial \phi_\mathrm{avg}$ is the partial derivative of $V_s$ with respect to $\phi_\mathrm{avg}$ and the $\sigma$ estimation is independent of SNL measurement since dividing $V_{n}$ with $V'_s$ can cancel $V_{sn}$ out. 

\begin{figure}[ht]
\centering
\includegraphics[width=0.65\linewidth]{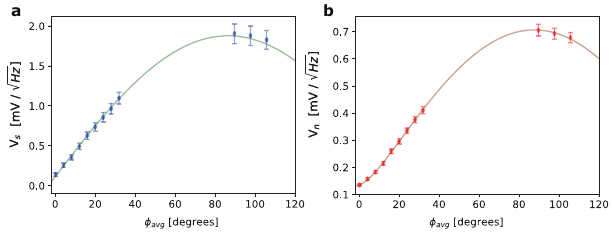}
\caption{\textbf{(a)} and \textbf{(b)}: Fitting from measured $V_s$ and $V_n$ in different $\phi_\mathrm{avg}$. With the fiting result, $\sigma$ is estimated by using Eq. (\ref{eq:min_sigma}).}
\label{fig:fitting}
\end{figure}

In the experiment we give an identical local phase shift to all 4 modes so that $\phi_j=\phi_\mathrm{avg}$ for all $j=1$ to $4$, and change the value of $\phi_\mathrm{avg}$ around both the squeezing $\phi_\mathrm{avg}=0$ and the anti-squeezing $|\phi_\mathrm{avg}|=90^\circ$. The $\phi_\mathrm{avg}$ we choose to induce as well as the fitting to Eq. (\ref{eq:VsVn}) with measured $V_{s}$ and $V_n$ from Figure \ref{fig6} is shown in Figure \ref{fig:fitting}. 
In the $V_s (\phi_\mathrm{avg})$ fitting, the parameters to fit are the slope $ k = \sqrt{\frac{2}{M}} V_{sn} \alpha_e $ and $\theta_1$. In the $V_n$ fitting, the squeezing noise voltage scaled by SNL, $k_{sq}= V_{sn} \cdot v_{sq}$, the anti-squeezing noise voltage scaled by SNL, $k_{asq} = V_{sn} \cdot v_{asq}$, and $\theta_2$ are fitting parameters. With the fitting result, we estimate the small angle sensitivity of our system $\sigma_{min}$ by 
\begin{equation}
\label{eq:min_sigma}
\sigma_{min} = \frac{V_n(\phi_\mathrm{avg}=0)}{V'_s(\phi_\mathrm{avg}=0)}.
\end{equation}
We do fitting for 5 different pump power of OPO and find the fitted values of $\theta_1$ and $\theta_2$ are $3.4 ^{\circ}\pm0.2^{\circ}$ and $1.6 ^{\circ}\pm0.6^{\circ}$, respectively. 
These values are reasonably small, and in principle could be further reduced by better locking and phase modulation techniques.
For the most sensitive case (maximum squeezing rate) in our result, the fitted $\theta_1$ and $\theta_2$ indicate $\sigma_{min}$ could have been further improved by $\sim0.2\%$ and $\sim0.9\%$, respectively. 
The $\sigma_{min}$ extracted in Eq. (\ref{eq:min_sigma}) is shown in our experiment results for the entangled approach in main text Fig 3 as $\sigma_e$. The uncertainty of $V'_s$ is obtained from the fitting, and the uncertainty of $V_n$ is obtained from the standard deviation of 2000 measurements. The error bars for $\sigma$ in Fig.~3 are calculated by error propagation of Eq. (\ref{eq:min_sigma}).

A similar analysis method is used for the separable approach, but the PSDs used in the separable approach are from only one HD instead of averaged HD outputs. By removing the BSN, our setup gives the separable approach of $M=1$. To compare with the entangled approach of $M=4$, we rescale our result with $1/\sqrt{M}$ as a result of classical averaging. The scaled sensitivity is quoted as our experiment result for the separable approach in main text Fig.~3 as $\sigma_s$.

\subsection{Resource counting}

In this section, we show how to experimentally measure the average photon number per mode that we use in the phase sensing. 

For the entangled approach, we estimate $M N_e= M N_{e,coh} + M N_{e,sqz}$ by comparing the joint measurement PSDs for squeezing and anti-squeezing quadrature to that for SNL, where $M=4$ is the mode number. With the notation defined above, the average number of squeezed photons for all modes in the entangled approach are obtained by comparing $V_n$ to $V_{sn}$ with
\begin{equation}
\label{Nesqz}
    M \cdot N_{e,sqz} = \frac{1}{2}\left( \langle \Delta \hat P_{avg}^2 \rangle +  \langle \Delta \hat X_{avg}^2 \rangle - 1\right) = \frac{1}{4}\left[\frac{V^2_n(\phi_\mathrm{avg} = 0^\circ)}{V^2_{sn}} + \frac{V^2_n(\phi_\mathrm{avg} = 90^\circ)}{V^2_{sn}}-2\right],
\end{equation}
Similarly, the average number of coherent photons are obtained by comparing $V_s$ to $V_{sn}$ with
\begin{equation}
\label{Necoh}
M \cdot N_{e,coh} = \eta \alpha_e^2 =\frac{V^2_s(\phi_\mathrm{avg}=90^\circ) }{ 4 V^2_{sn}}.
\end{equation} 
With Eq. (\ref{Nesqz}) and (\ref{Necoh}), $N_e= N_{e,coh} + N_{e,sqz}$ gives the $N$ values in main text Fig.~3 for the entangled approach $\sigma_{e}$. \changed{The error bars for $N$ in Fig.~3, entangled approach are calculated by error propagation of Eqs. (\ref{Nesqz}-\ref{Necoh}).}

For the separable approach, we use a very similar technique. However, the PSD is from a single HD instead of joint measurement. Explicitly, the photon number per mode for the separable approach is $N_s= N_{s,coh} + N_{s,sqz}$, with
\begin{equation}
    N_{s,sqz} = \frac{1}{2}\left( \langle \Delta \hat p_{j}^2 \rangle +  \langle \Delta \hat x_{j}^2 \rangle - 1\right) = \frac{1}{4}\left[\frac{V^2_n(\phi = 0^\circ)}{V^2_{sn'}} + \frac{V^2_n(\phi = 90^\circ)}{V^2_{sn'}}-2\right]
\end{equation}
and
\begin{equation}
   N_{s,coh} = \frac{ V^2_s(\phi=90^\circ) }{4 V^2_{sn'}},
\end{equation} 
where $\phi$ is the phase shift of the single mode, and we use $V_{sn'}$ to denote the 1-mode SNL, which is about $1/4$ of the 4-mode SNL $V_{sn}$ used in the entangled approach. The photon number per mode $N_s= N_{s,coh} + N_{s,sqz}$ gives the $N$ values in main text Fig.~3 for the separable approach $\sigma_{s}$. The error bars for $N$ in Fig.~3, separable approach are calculated by error propagation of Eqs. (60-61).

\end{document}